\documentclass[onecolumn]{emulateapj}
\usepackage{apjfonts}

\shorttitle{Braking the Gas in $\beta$~Pic}
\shortauthors{Fern\'andez, Brandeker, \& Wu}
\begin{document}
\title{Braking the gas in the $\beta$~Pictoris disk}

\author{Rodrigo Fern\'andez\altaffilmark{1}, Alexis Brandeker, and Yanqin Wu}
\affil{Department of Astronomy and Astrophysics, University of Toronto. 60 St.~George Street, Toronto, 
ON M5S\,3H8, Canada}

\altaffiltext{1}{e-mail: fernandez@astro.utoronto.ca}
\slugcomment{Received 2005 September 8; Accepted 2005 December 28}
\begin{abstract}
Metallic gas detected in the $\beta$~Pictoris circumstellar debris disk
raises many questions. The origin of this gas is unclear and its
very presence is difficult to explain: many constituents of the gas
are expected to be radiatively accelerated outward, yet their motion
appears to be consistent with Keplerian rotation out to at least
$300$~AU. Hydrogen has previously been hypothesized to exist in the
disk, acting as a braking agent, but the amount required
to brake individual elements conflicted with observed upper limits.

To resolve this discrepancy, we search for alternative braking
mechanisms for the metallic gas. We find that all species affected by
radiation force are heavily ionized. Frequent Coulomb collisions
couple the ions into a single fluid, reducing the radiation force on
species feeling the strongest acceleration. For a gas of solar
composition, the resulting total radiation force still exceeds
gravity, while a gas of enhanced carbon abundance could be
self-braking.  We also explore two other braking agents: collisions
with dust grains and neutral gas. Grains surrounding $\beta$\,Pic are
photoelectrically charged to a positive electrostatic potential. If a
significant fraction of the dust grains are carbonaceous
($10\%$ in the midplane and larger at higher altitudes), 
ions can be slowed down to satisfy the observed velocity constraints. In this
case, both the gas kinematics and spatial distributions are expected
to coincide with those of small grains, the latter being indeed
observed. For neutral gas to brake the coupled ion fluid, we find the
minimum required mass to be $\approx 0.03$\,M$_{\earth}$, consistent
with observed upper limits of the hydrogen column density, and
substantially reduced relative to previous estimates.

Our results favor a scenario in which metallic gas is generated by
grain evaporation in the disk, perhaps during grain-grain
collisions. We exclude a primordial origin for the gas, but cannot
rule out the possibility of its production by
falling evaporating bodies near the star. We discuss the implications
of this work for observations of gas in other debris disks.
\end{abstract}


\section{INTRODUCTION}
\label{s:intro}

The evolution of circumstellar disks is intimately connected to the
process of star and planet formation. In the standard isolated
low-mass star formation scenario (e.g., \citealt*{shuadamslizano}),
disks begin their existence as \emph{protostellar} accretion disks,
later turning into \emph{protoplanetary} disks, once the newborn star
reaches the T~Tauri phase. The late evolutionary stages of
circumstellar disks are not well understood. Once the star has reached
the main sequence and the planet formation process is complete, the
disk is thought to have been depleted of its primordial gas, keeping
small amounts generated by either stellar winds or evaporation of
solid bodies (e.g., \citealt{lagrange2000}). In the few cases where
late disks are observable, lifetimes of small dust grains ($\lesssim$
1\,$\mu$m) are shorter than the estimated ages of those systems,
implying a dust replenishment mechanism \citep{backman93}. These kind
of late circumstellar disks are usually referred to as
\emph{debris disks}. The gas lifetime in those systems, currently unknown, determines
the maximum period over which protoplanetary cores can accrete gas and form gaseous planets 
(e.g., \citealt{bodenheimer_lin}), and is related the formation of terrestrial planets and cores of giant 
planets by the increased efficiency of planetesimal accretion due to the presence of gas \citep{rafikov04}.

So far, the best studied debris disk is that of $\beta$~Pictoris, a
nearby (19.3 $\pm$ 0.2 pc, \citealt{crifo97}), A5V star. The dust
component of the disk has been extensively studied since it revealed
itself as infrared excess in the stellar spectrum
\citep{aumann85}. The favorable edge-on orientation has allowed
direct imaging in thermal emission
\citep{lagage94}, as well as in scattered light (\citealt{smith84}, \citealt{heap00}), 
indicating a disk size $\sim$1\,000\,AU. 
The gas component has been more elusive. \citet{freudling95} set a
$3\sigma$ upper limit on the column density of H\,I,
$N_\mathrm{H\,I}\leq 10^{19}$\,g\,cm$^{-2}$, by non-detection of the
21\,cm line, while \citet{thi01} and \citet{lecavelier01} searched
for signatures of H$_\mathrm{2}$ in the IR and UV, respectively,
obtaining discrepant results: 50\,M$_\earth$ in the first case
(corresponding to N$_{\mathrm{H}_2}\sim 3\times 10^{21}$\,cm$^{-2}$),
and an upper limit N$_{\mathrm{H}_2}\leq 10^{18}$\,cm$^{-2}$ 
($\sim$0.1\,M$_\earth$) in the second. Metallic elements have been detected
through stable and variable circumstellar lines
\citep{hobbs85, ferlet87}. The variable, redshifted lines have been attributed to falling evaporating bodies (FEBs) 
in highly eccentric orbits \citep{ferlet87}, whereas the stable
component has no well established explanation for its origin.
\citet{lagrange98} proposed three possible gas formation scenarios: stellar wind, star-grazing comet evaporation, and dust grain 
evaporation near the star. Recent improvements in the gas
characterization include spatially resolved emission lines from stable
metallic species, which revealed them to be consistent with Keplerian
rotation \citep{olofsson01}, and spatially extended out to at least
$300$\,AU from the star \citep{brandeker04}.

It was realized early on (e.g., \citealt{beust89}) that certain
elements in the gas disk of $\beta$\,Pic should be subject to a very
strong radiation force from the star, overcoming gravitational
attraction. However, besides Keplerian rotation, stable lines do not
show any substantial radial velocities relative to the star (e.g.,
\citealt{lagrange98}, \citealt{brandeker04}). This implies the
presence of a braking mechanism acting in the disk.

The first model of how gas might be braked and produce the stable
absorption lines was presented by \citet{lagrange98}. They proposed a
scenario in which gas is injected into the disk due to FEBs. Since
neutral hydrogen is not affected by the stellar radiation force, it would
accumulate in the region of gas generation, in the form of an annulus
located a few AU from the star. The elements affected by radiation
force would then be slowed down temporarily in this annular region,
producing the observed stable absorption features. After passing this
hydrogen-rich region, elements would then be accelerated again,
leaving the system at high velocities. This model reproduces the
stable absorption lines, and is consistent with the FEBs scenario for
the variable lines. However, the fact that elements affected by
radiation force reach high velocities after passage by the annulus,
which extends only out to a few AU from the star, is incompatible with
the emission from $\sim$300\,AU. \citet{brandeker04} calculated the
amount of material necessary to slow down the most strongly
accelerated particles to within observational constraints by means of
neutral-neutral collisions, finding that $\sim$50\,M$_\earth$ of
hydrogen are needed. According to \citet{thebault05}, there could be
no more than 0.4\,M$_\earth$ of gas in the disk, as it would affect
the dynamics of the dust to the degree that dust production models
become incompatible with observations.

As an alternative braking mechanism, \citet{alexis_thesis} explored
the slowing down of ionized species by a large-scale magnetic field,
concluding that only a toroidal field could remove the radial velocity
components of ions. \citet{alexis_thesis} showed that unless the field
is stronger than $\sim$1\,mG at 100\,AU, this configuration would be
unstable to currents generated by radiation force. He also raised the
question of the origin of such an ad-hoc field geometry.

In this paper we address the braking of the stable metallic gas in the
entire $\beta$\,Pic disk. Given the spectral type of the star and the
fact that the disk is optically thin, we expect several gaseous
species to be ionized. This prompts us to explore the role of Coulomb
collisions among ionized gas particles, which may be an important
braking agent due to the long range of electromagnetic
interactions. We proceed then to explore the interaction of this
ionized gas with dust grains, which are photoelectrically
charged. Finally, we extend the study of \citet{lagrange98} to explore
whether ion-neutral collisions are able to brake particles affected by
radiation force far out in the disk, and the amount of neutral gas
required to satisfy the observed velocity constraints.

The paper is organized as follows: in \S\ref{s:stage} we explore the physical
conditions of the gas, and the constraints on the possible braking
mechanisms that this imposes. In \S\ref{s:brake} we study ion-ion collisions,
ion-grain collisions, and revisit ion-neutral interactions. In \S\ref{s:discuss} we
discuss the implications of our results for gas generation and
extend it to similar systems. Our conclusions follow in \S\ref{s:conclusions}.

\section{SETTING THE STAGE FOR BRAKING}
\label{s:stage}

In this section we explore the physical environment of the gaseous
disk. This will enable us to characterize the requirements the
braking mechanisms have to satisfy.

\subsection{Radiation Force}
\label{s:radfor}

The radiation force $F_\mathrm{rad}$ acting on a given atom originates by a net momentum transfer due to scattering events,
\begin{equation}
F_\mathrm{rad} = \frac{1}{c}\sum_{j<k} F_\lambda\sigma_{jk},
\end{equation}
where $c$ is the speed of light, $\sigma_{jk}$ is the cross section
for the transition between atomic levels $j$ and $k$ integrated over
wavelength, and $F_\lambda$ is the stellar flux per unit wavelength
evaluated at the line center. Since the stellar flux decays like the
inverse square of the distance from the source, the ratio $\beta$
between radiation and gravitational forces is distance-independent:
\begin{equation}
\label{eq:beta_def}
\beta \equiv \frac{F_\mathrm{rad}}{F_\mathrm{grav}} = \frac{1}{8\pi c^2}\frac{r^2}{G M m}\sum_{j<k} \frac{g_k}{g_j} A_{kj} \lambda_{jk}^4 
F_\lambda,
\end{equation}
where we have expressed $\sigma_{jk}$ in terms of known atomic
quantities (e.g., \citealt{hilborn82}). Here, $g_j$ and $g_k$ are the
statistical weights of levels $j$ and $k$, $A_{kj}$ is the Einstein
coefficient for spontaneous emission from level $k$ to $j$,
$\lambda_{jk}$ the wavelength of the corresponding transition, $G$ is
the gravitational constant, $M$ the stellar mass, and $m$ the mass of
the atomic species under consideration. 

The energy of a particle of mass $m$ in a bound orbit with semi-major axis
$a$ around an object of mass $M$ is $E_{\mathrm{bind}} = -GMm/(2a)$. When the 
contribution of 
radiation is added, the energy of the system increases by an amount
$E_{\mathrm{rad}} = \beta GMm/r$, where $r$ is the distance to the star at a given
point in the orbit. For a circular orbit ($r=a$), the particle ceases to
be bound if $\beta > 0.5$. If $\beta<0.5$ the particle remains in a bound
orbit, but ``feeling" a lower dynamical mass $(1-\beta)M$. 

Values for $\beta$ in this system are available in the literature for
some elements, but we find it necessary to calculate a set of $\beta$
values for as many elements as possible. The low temperatures and
densities of the gas in the disk (at distances $\gtrsim$ few AU from
the star, see \S\ref{s:ionstate}) imply that radiative de-excitations are faster
than either collisions or radiative excitations, thus most atoms are
in the ground state. This implies $j=0$ in
equation~(\ref{eq:beta_def}), simplifying calculations considerably.
Since we do not expect species to be more than twice ionized (see
\S\ref{s:ionstate}), we study the first three ionization states of elements from
H to Ni. The atomic data are taken from the NIST Atomic Spectra
Database version
2.0~\citep{asd}\footnote{http://physics.nist.gov/asd2}. For elements
with multiple ground states (e.g., Fe\,I), we assume that the
population of each level is proportional to its statistical weight
(e.g., \citealt{lagrange96}). The uncertainties in $A_{k0}$ range from
$<$3\% for the strongest lines to $\sim$100\% for the weakest.

The stellar flux is approximated with a PHOENIX model spectrum
\citep*{hauschildt99} for a star of effective temperature
$T_\mathrm{eff}=8\,000$\,K and surface gravity $\log{g} = 4.2$. Given
that $\beta$\,Pic is rotating at $130\pm 4$\,km\,s$^{-1}$
\citep{royer02}, the spectral features are significantly broadened. A
proper rotation of the spectrum requires detailed resolution of the
photospheric absorption profiles, thus a special version of the model
atmosphere was used in this work, which has constant spacing in
wavelength, namely 0.05\,\AA\ in the region 1\,000--10\,000\,\AA, and
0.5\,\AA\ in the region 10\,000--50\,000\,\AA\ (P.~Hauschildt \&
I.~Kamp, private communication). A rotational broadening of this model
spectrum was performed (e.g., \citealt{gray76}), assuming a linear
limb darkening law with $\varepsilon=0.5$. This choice adds an
uncertainty in the relative flux level at the bottom of photospheric
lines ranging from 2\% in the visual to 10\% in the UV. The spectrum
was then flux calibrated using photometric information from the
Tycho-2 Catalog
\citep{hog00} and the distance to the star. We estimate the error of 
this flux calibration to be $\sim$4\%.

Table~\ref{t:betavalues} shows the results of our calculations for
elements from H to Ni. The mass of the star used is $M =
1.75$\,M$_\sun$ \citep{crifo97}. Calculation of uncertainties is
explained in Appendix~\ref{s:uncertainty}. The agreement level with previously
calculated values is highly variable. As an example,
\citet{lagrange96} calculated $\beta$ values using calibrated HST
spectra. Our results agree with their values for Fe\,II, Al\,II and
Al\,III within our uncertainties, but their result for Mg\,II differs
from ours by a factor of 2. We take the good agreement between their
Fe\,II value and ours as a consistency check, given the great number
of transitions required to calculate $\beta$ for this ion.
Nevertheless, when looking at our results, it should be kept in mind
that they were obtained with a generic model spectrum, therefore
subject to possible systematic deviations from the particular
$\beta$\,Pic case.
\begin{deluxetable*}{lclclclc}
\tablecaption{Radiation force coefficients obtained from rotationally broadened $\beta$\,Pic-like spectrum\label{t:betavalues}}
\tablewidth{0pt}
\tablehead{\colhead{Ion} & \colhead{$\beta$\tablenotemark{a,b}} & \colhead{Ion} & \colhead{$\beta$\tablenotemark{b}} & 
\colhead{Ion} & \colhead{$\beta$\tablenotemark{a,b}} & \colhead{Ion} & \colhead{$\beta$\tablenotemark{a,b}}}
\startdata
H\,I    &  $(1.6\pm 0.1)10^{-3}$    &  F\,I    &  $0$                    &  S\,I       & $0.56\pm 0.09$         & V\,I     &  $72\pm 4$  \\   
He\,I   &  $0$                      &  F\,II   &  $(3.5\pm 0.9)10^{-6}$  &  S\,II      & $(9.0\pm 1.0)10^{-5}$  & V\,II    &  $4.4\pm 0.2$\\  
He\,II  &  \nodata                  &  F\,III   & $(5.0\pm 1.0)10^{-9}$  &  S\,III     & $(2.0\pm 1.0)10^{-4}$  & V\,III   &  $0$       \\
Li\,I   &  $900\pm 40$              &  Ne\,I    & $0$                    &  Cl\,I      & $(2.3\pm 0.4)10^{-3}$  & Cr\,I    &  $93\pm 5$  \\  
Li\,II  &  $0$                      &  Ne\,II   & $0$                    &  Cl\,II     & $(3.7\pm 0.4)10^{-7}$  & Cr\,II   &  $(6.0\pm 3.0)10^{-7}$  \\   
Li\,III & \nodata                   &  Ne\,III  & $(9.0\pm 2.0)10^{-8}$  &  Cl\,III    & $(3.0\pm 2.0)10^{-6}$  & Cr\,III  &  \nodata \\  
Be\,I   &  $62\pm 7$                &  Na\,I    & $360\pm 20$            &  Ar\,I      & $(1.7\pm 0.3)10^{-6}$  & Mn\,I    &  $28\pm 3$  \\
Be\,II  &  $124\pm 6$               &  Na\,II   & $0$                    &  Ar\,II     & $0$                    & Mn\,II   &  $7\pm 1$  \\
Be\,III &  $0$                      &  Na\,III  & $0$                    &  Ar\,III    & $(1.5\pm 0.2)10^{-7}$  & Mn\,III  &  \nodata \\
B\,I    &  $30\pm 10$               &  Mg\,I    & $74\pm 8$              &  K\,I       & $200\pm 20$            & Fe\,I    &  $27\pm 2$  \\
B\,II   &  $0.07\pm 0.04$           &  Mg\,II   & $9\pm 2$               &  K\,II      & \nodata                & Fe\,II   &  $5.0\pm 0.3$  \\  
B\,III  &  $19\pm 1$                &  Mg\,III  & $0$                    &  K\,III     & $(4.4\pm 0.2)10^{-4}$  & Fe\,III  &  $(3.0\pm 0.6)10^{-7}$  \\  
C\,I    &  $(3.3\pm 0.1)10^{-2}$    &  Al\,I    & $53\pm 6$              &  Ca\,I      & $330\pm 40$            & Co\,I    &  $16\pm 1$  \\  
C\,II   &  $(2.3\pm 0.2)10^{-3}$    &  Al\,II   & $0.36\pm 0.05$         &  Ca\,II     & $50\pm 10$             & Co\,II   &  $0$       \\  
C\,III  &  $(8.5\pm 0.9)10^{-6}$    &  Al\,III  & $12\pm 1$              &  Ca\,III    & \nodata                & Co\,III  &  $(4.0\pm 2)10^{-7}$  \\  
N\,I    &  $(2.1\pm 0.1)10^{-4}$    &  Si\,I    & $6.0\pm 0.6$           &  Sc\,I      & $220\pm 20$            & Ni\,I    &  $26\pm 2$  \\  
N\,II   &  $(7.5\pm 0.5)10^{-6}$    &  Si\,II   & $9\pm 9$               &  Sc\,II     & $(1.3\pm 0.4)10^3$     & Ni\,II   &  $(7.0\pm 2.0)10^{-2}$  \\  
N\,III  &  $(7.0\pm 1.0)10^{-6}$    &  Si\,III  & $(5.8\pm 0.6)10^{-4}$  &  Sc\,III    & $(9.0\pm 3.0)10^{-2}$  & Ni\,III  &  $(3.0\pm 2.0)10^{-7}$  \\  
O\,I    &  $(3.3\pm 0.2)10^{-4}$    &  P\,I     & $3.4\pm 0.6$           &  Ti\,I      & $97\pm 5$              & p\tablenotemark{c}& $4.4\times 10^{-11}$  \\
O\,II   &  $(3.1\pm 0.7)10^{-9}$    &  P\,II    & $(2.2\pm 0.3)10^{-3}$  &  Ti\,II     & $28\pm 2$              & e\tablenotemark{c}& $0.27$\\   
O\,III  &  $(6.5\pm 0.6)10^{-7}$    &  P\,III   & $(5.0\pm 2.0)10^{-4}$  &  Ti\,III    & $(5.0\pm 0.1)10^{-4}$  & \nodata  & \nodata\\
\enddata
\tablenotetext{a}{$\beta=0$ means that no ground state transitions are known in the range 1\,000\,\AA\ $< \lambda <$ 50\,000\,\AA}
\tablenotetext{b}{Empty entries mean that no atomic data are available}
\tablenotetext{c}{Proton (H\,II) and electron values are calculated using Thomson cross section}
\end{deluxetable*}

\subsection{Ionization State of the Gas}
\label{s:ionstate}

Since we are interested in relating the dynamics of the gas to
electrostatic interactions, we need to analyze in detail the
ionization state of the gas. Given that the disk is optically thin,
and the densities are low, the ionization state is determined by
photoionization and radiative recombination. The temperature of the
gas is below 300\,K \citep{kamp01}, thus recombination rates are not
very sensitive to the exact value of the temperature, depending most
strongly on the electron density. Given the spectral type of the star
(A5V), we expect most elements to be either singly ionized, or
neutral.

We thus wrote a simple photoionization code which calculates the
densities of neutral and ionized elements from H to Ni for an
optically thin gas disk, with the corresponding electron density, and
assuming solar composition. The latter choice is motivated by
observations showing refractory elements having solar abundance
relative to each other \citep{lagrange95}. The stellar ionizing
radiation is obtained from our rotated and flux-calibrated model
spectrum, whereas interstellar background UV field and and cosmic ray 
ionization rate
($2\times10^{-17}$\,s$^{-1}$\,atom$^{-1}$) were taken from
\citet{weingartner01} and \citet{spitzer78}, respectively. Radiative
recombination coefficients and photoionization cross sections were
taken from the Cloudy C96.01 atomic data library
\citep{ferland98}. Recombination of ions on dust grain surfaces, which 
can be the dominant recombination process in the interstellar medium,
is not included as the dust in the disk is expected to
have a \textit{highly positive} charge (Appendix~\ref{s:charge}), 
in contrast to the negatively charged grains of the interstellar medium.

\begin{figure}
\plotone{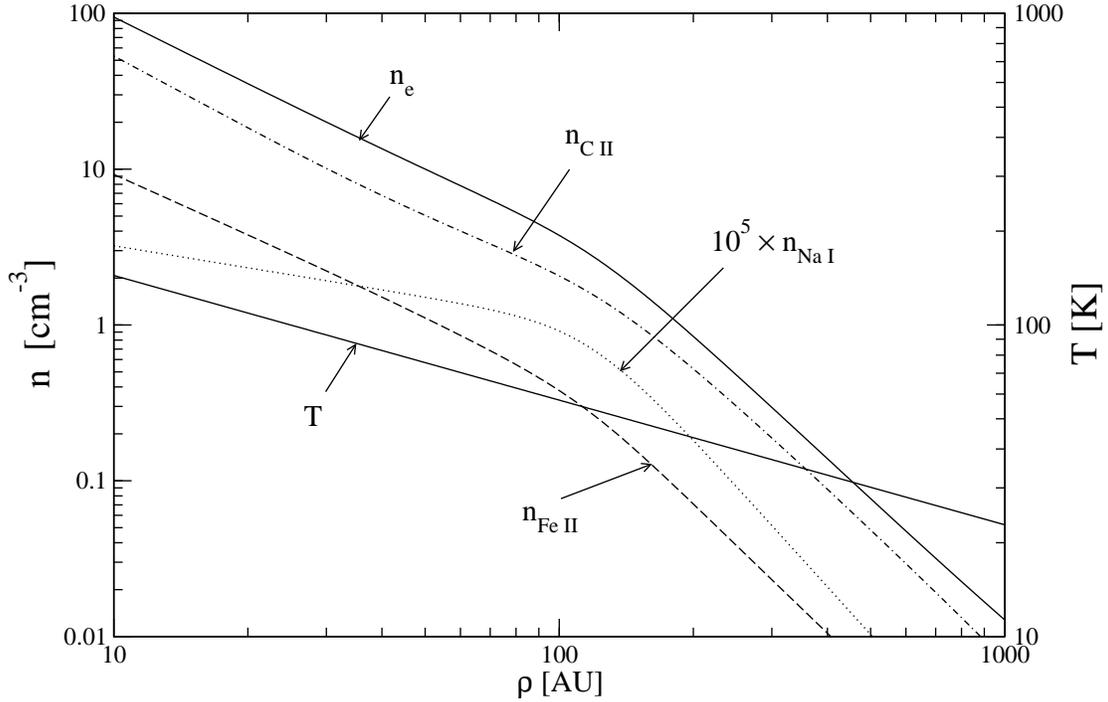}
\caption{Plotted against cylindrical radius $\rho$ are: average midplane dust temperature $T$ (eq.~\ref{eq:T_mean}, right axis), which we take
to be the same as the local gas temperature; and midplane number
densities (left axis) of electrons and selected species. Na\,I is
obtained from observations \citep{brandeker04}, the remaining species
are obtained from our photoionization code. C is partially ionized
($\sim$50\%) throughout the disk (and the main contributor of electrons), while Na and Fe are almost
completely ionized.}
\label{fig:Tne}
\end{figure}

At low temperatures the chemistry of the gas is not affected
significantly if one sets $T_\mathrm{gas} = T_\mathrm{dust}$
\citep{kamp01}, where $T_\mathrm{gas}$ 
is the gas temperature and $T_\mathrm{dust}$ is the dust
temperature. We thus adopt a temperature profile corresponding to
grains of size $a$ (e.g., \citealt{kamp01})
\begin{equation}
\label{eq:kamp_temp_profile}
T_\mathrm{a} = 282.5\left( \frac{L}{\textrm{L}_\sun}\right)^{1/5}\left(\frac{\mathrm{AU}}{r}\right)^{2/5}\left(\frac{\mu \mathrm{m}}{a}\right)^{1/5}\textrm{\,K},
\end{equation}
where $L$ is the stellar luminosity ($11$\,L$_\sun$ for $\beta$\,Pic), and $r$ is the distance to the star. To
eliminate the dependence on grain size, we take a mean temperature weighted by grain surface area and size distribution
\begin{equation}
\label{eq:T_mean}
T_\mathrm{gas}^4 = \langle T^4_\mathrm{a}\rangle = \frac{\int T^4_\mathrm{a} a^2 dn_\mathrm{a}}{\int a^2 dn_\mathrm{a}}.
\end{equation} 
We adopt $dn_\mathrm{a} \propto a^{-3.5}$ and perform the integral
over the range $a_\mathrm{min} < a < a_\mathrm{max}$, with
$a_\mathrm{min} = 1$\,$\mu$m and $a_\mathrm{max} = 1$\,km. The chosen
size distribution implies that the total grain surface area, and
therefore our weighted temperature, is dominated by values
corresponding to the smallest grains, being insensitive to the choice of $a_\mathrm{max}$. 
The resulting temperature profile in the disk midplane is shown in Figure~\ref{fig:Tne} (right
axis). 

The observational constraint on the gas density comes from the fit to Na\,I emission \citep{brandeker04}, which has the form
\begin{equation}
\label{eq:sodium_profile}
n_\mathrm{Na\,I}(\rho,z) = 10^{-5}\left[\left(\frac{\rho}{\rho_0}\right)^{0.94} + \left(\frac{\rho}{\rho_0}\right)^{6.32}\right]^{-1/2}\exp{\left[ -\left(\frac{z}{0.17 \rho}\right)^2\right]}\textrm{\,cm}^{-3}, 
\end{equation}
where $\rho$ is the distance to the star in the disk midplane, $z$ the
height above the midplane ($r^2 = \rho^2 + z^2$), and $\rho_0 =
117$\,AU. The Na\,I profile in the disk midplane is shown in
Figure~\ref{fig:Tne}, scaled by a factor $10^5$. Our photoionization
code numerically solved for the densities of neutral and ionized
species that are consistent with this Na\,I profile, the temperature
profile of equation~(\ref{eq:T_mean}), and our assumption of solar
composition. The sensitivity of our results to the adopted temperature profile
is discussed in \S\ref{s:tempsens}.

Our results show that H, He, N, O, F, Ne, and Ar, if present, are
almost completely neutral, their marginal ionization ($\sim$10$^{-6}$)
being determined by interstellar UV flux and cosmic rays. This group
is unaffected by radiation force in its neutral phase. The elements C,
Cl, Be, P, and S are partially ionized, with ionization fractions 
ranging from $\sim$50\% for carbon to $\gtrsim$95\% for beryllium. Inside 
$\sim$50\,AU the stellar ionization dominates, whereas
UV background takes over at larger distances. While neutral carbon, chlorine,
and sulfur feel little or no radiation pressure, the neutral phases of
phosphorus and beryllium are accelerated. The potential of these species
to trace the velocity of neutrals will be discussed below.
The remaining species (including Na) are ionized
to fractions $\gtrsim$99.9\%. Electron densities in the disk midplane
are shown in Figure~\ref{fig:Tne}, together with density profiles of
C\,II and Fe\,II, which are the most abundant species by mass in the
partially-, and heavily-ionized group.

The group with ionization $\gtrsim$99.9\% contains all species
affected strongly by radiation force. This is crucial for explaining
the dynamics of the gas. From Table~\ref{t:betavalues} we see that for
most of these species, the highest $\beta$ corresponds to the neutral
phase (exceptions being Be and Sc), which has a very short lifetime
compared to that of the ionized phase. Once ionized, elements feel a
lower radiation force and can undergo Coulomb interactions with other
charged particles and lose the excess momentum acquired from photons
(\S\ref{s:ioncoll}).

A quantity of relevance is the velocity boost a neutral particle
acquires before being ionized, $v_\mathrm{ion}$. In the optically thin
case, the ratio between the number per unit time of scattering photons
to that of ionizing stellar photons depends only on the shape of the
stellar spectrum, being \emph{independent} of the distance to the
star. The number of ionizing stellar photons hitting an atom per unit
time is the stellar ionization rate
\begin{equation}
\Gamma_* = \int_0^{\infty} \frac{F_\lambda}{(hc/\lambda)} \sigma_\mathrm{ion} d\lambda,
\end{equation}
where $\sigma_\mathrm{ion}$ is the ionization cross section, the ionization time (or lifetime as neutral) being the inverse of 
$\Gamma_*$. The velocity boost before ionization is then 
\begin{equation}
\label{eq:v_ion}
v_\mathrm{ion} \simeq \beta\frac{GM}{r^2}\frac{1}{\Gamma_*},
\end{equation} 
where $\Gamma_*$ is evaluated at distance $r$ from the
star. Table~\ref{t:beta_velocities} shows the results for elements
with $\beta > 0.5$, for which the ionization rate due to interstellar
UV is negligible relative to $\Gamma_*$. The most relevant feature is
that, with the exception of Be\,I and P\,I, all neutrals feeling
radiation force achieve velocities below 3.5\,km\,s$^{-1}$ before being
ionized. Since the acceleration in the neutral phase is nearly
constant, the mean radial velocity is $v_\mathrm{ion}/2$, provided no
braking mechanism acts on particles in the neutral phase. This is
consistent with observed values for neutrals: $v_\mathrm{Na\,I} =
1.2\pm 0.3$\,km\,s$^{-1}$, $v_\mathrm{Fe\,I} = 0\pm
0.3$\,km\,s$^{-1}$, and $v_\mathrm{Ni\,I} = 0.4\pm
0.4$\,km\,s$^{-1}$\citep{brandeker04}. Hence, as long as ionized
particles are effectively braked, there is no need to find a braking
mechanism for high-$\beta$ neutrals. Observed radial velocities of
Be\,I and P\,I would be able to test these
predictions. Unfortunately, their strongest ground state transitions
are in the UV, which requires observations from space. The expected
line equivalent widths, derived using our ionization model, would be
10\,m\AA\ for P\,I $\lambda$1774, while Be\,I $\lambda$2350 would be
hardly detectable at 3\,$\mu$\AA.

\begin{deluxetable}{lccc}
\tablecaption{Velocities before ionization for neutral species with $\beta > 0.5$\label{t:beta_velocities}}
\tablewidth{0pt}
\tablehead{
\colhead{Element} & \colhead{$\Gamma_\mathrm{star}$\tablenotemark{a}} & \colhead{$v_{\mathrm{ion}}$\tablenotemark{b}} & \colhead{$n_\mathrm{I}/n_\mathrm{II+}$\tablenotemark{c}}\\ 
                  & \colhead{(s$^{-1}$)}                              & \colhead{(km\,s$^{-1}$)}                       &     
}
\startdata
Li\,I                   & $9.0\times 10^{-6}$  &  $0.11$              & $4.4\times 10^{-6}$ \\
Be\,I                   & $1.1\times 10^{-9}$  &  $58$                & $4.4\times 10^{-2}$ \\
B\,I                    & $7.5\times 10^{-8}$  &  $0.41$              & $6.1\times 10^{-4}$ \\
Na\,I\tablenotemark{d}  & $1.1\times 10^{-7}$  &  $3.3$               & $3.2\times 10^{-4}$ \\
Mg\,I                   & $6.8\times 10^{-8}$  &  $1.1$               & $6.4\times 10^{-4}$ \\
Al\,I                   & $1.1\times 10^{-4}$  &  $5.0\times 10^{-4}$ & $8.5\times 10^{-7}$ \\
Si\,I                   & $3.9\times 10^{-7}$  &  $0.02$              & $1.3\times 10^{-4}$ \\
P\,I                    & $4.0\times 10^{-10}$ &  $8.7$               & $1.5\times 10^{-1}$ \\
S\,I                    & $4.0\times 10^{-10}$ &  $1.5$               & $6.4\times 10^{-2}$ \\
K\,I                    & $4.4\times 10^{-7}$  &  $0.5$               & $1.5\times 10^{-4}$ \\
Ca\,I                   & $1.3\times 10^{-5}$  &  $0.03$              & $3.5\times 10^{-6}$ \\
Sc\,I                   & $6.4\times 10^{-8}$  &  $3.5$               & $7.0\times 10^{-4}$ \\
Ti\,I                   & $1.5\times 10^{-7}$  &  $0.6$               & $3.0\times 10^{-4}$ \\
V\,I                    & $4.0\times 10^{-7}$  &  $0.2$               & $1.2\times 10^{-4}$ \\
Cr\,I                   & $1.0\times 10^{-7}$  &  $0.9$               & $4.8\times 10^{-4}$ \\
Mn\,I                   & $6.8\times 10^{-8}$  &  $0.4$               & $7.6\times 10^{-4}$ \\
Fe\,I\tablenotemark{d}  & $5.8\times 10^{-8}$  &  $0.5$               & $9.0\times 10^{-4}$ \\
Co\,I                   & $2.0\times 10^{-8}$  &  $0.8$               & $2.9\times 10^{-3}$ \\
Ni\,I\tablenotemark{d}  & $1.0\times 10^{-7}$  &  $0.3$               & $4.8\times 10^{-4}$ \\
\enddata 
\tablenotetext{a}{Stellar ionization rate at 100\,AU from the star.}
\tablenotetext{b}{Terminal velocity before ionization, \emph{independent} of distance.}
\tablenotetext{c}{Ratio of neutral (I) to ionized phase (II or higher) at $r = 100$\,AU in the disk midplane.}
\tablenotetext{d}{Species with measured radial velocity by \citet{brandeker04} [see \S\ref{s:ionstate} for 
values and uncertainties]}
\end{deluxetable}

\section{BRAKING THE IONIC FLUID}
\label{s:brake} 

We have established in the last section that the braking of neutral
species is not required, only ions need to be slowed down. This
changes the nature of the puzzle and makes it solvable. In this
section we explore collisional processes involving electrostatic
interactions that can account for the braking of ionized
particles. From simple dynamical considerations it is clear that any
braking medium needs to satisfy two requirements. First, a \emph{high
collision frequency} with ions, so that momentum can be effectively
exchanged with particles accelerated by radiation. Second, the medium
must provide a \emph{high inertia}, so that it is not dragged
along. We thus focus on three processes likely to operate in the disk:
ion-ion, ion-dust, and ion-neutral collisions.

\subsection{Collisions between Ions}
\label{s:ioncoll}

The fact that all the species affected by radiation pressure are
highly ionized prompted us to explore first the role of Coulomb
collisions among them. For treating the problem, we assume that
magnetic fields are weak enough in the regions of interest,
$\sim$100\,AU from the star, so that the dynamics of charged particles
is not influenced by them. Otherwise, complicated MHD effects may come
into play. The statistical treatment of collisions among charged
particles was studied by \citet{spitzer56}, based on results of
\citet{chandra41, chandra43} for gravitational interactions.
In Spitzer's framework, particles are categorized as \emph{test} and
\emph{field} particles, the former having an initial velocity relative
to the center of mass of the latter. Field particles are assumed to
follow a Maxwellian velocity distribution. Braking of test particles
is then characterized by the time it takes for the beam of test
particles to diffuse in velocity space.  This time corresponds to
\emph{dynamical friction} in the gravitational case, and is given by
\citet{spitzer56} as
\begin{equation}
\label{eq:t_S}
t_\mathrm{S}(\mathrm{t,f}) = \frac{m_\mathrm{t}^2 (k_\mathrm{B} T_\mathrm{f}) v}{8\pi n_\mathrm{f} Z_\mathrm{f}^2 Z_\mathrm{t}^2 e^4 (m_\mathrm{t} + m_\mathrm{f}) H(\xi)\ln{\Lambda}},
\end{equation} 
where the subscripts t and f correspond to test and field particles,
respectively. Here, $v$ is the velocity of test particles relative to
the center of mass of field particles, $m$ is the mass of individual
particles, $n$ is the number density, $Z$ is the charge in units of
the electron charge $e$ (throughout this work we assume
$Z_\mathrm{t}=Z_\mathrm{f} = 1$ for ions), $T_\mathrm{f}$ is the
temperature of the field particles, and $k_\mathrm{B}$ is Boltzmann's
constant. The quantities $\xi$ and $\Lambda$ are defined by
\begin{eqnarray}
\label{eq:xi}
\xi & = & \frac{v}{c_\mathrm{th,f}} \\
\label{eq:c_thermal}
c_\mathrm{th,f} & = & \sqrt{\frac{2k_\mathrm{B}T_\mathrm{f}}{m_\mathrm{f}}}\\ 
\label{eq:Lambda}
\Lambda & = & \frac{\lambda_\mathrm{D}}{b_{\mathrm{min}}} = \frac{3}{2Z_\mathrm{f} Z_\mathrm{t} e^3}\left( \frac{k_\mathrm{B}^3 T_\mathrm{e}^3}{\pi n_\mathrm{e}}\right)^{1/2},
\end{eqnarray} 
where $c_\mathrm{th,f}$ is the thermal velocity of field particles,
$\lambda_{\mathrm{D}}$ is the electron Debye screening length,
$b_{\mathrm{min}}$ is the minimum impact parameter for Coulomb
collisions, and $T_\mathrm{e}$ is the electron temperature. In this
context, electrons and ions are sufficiently thermalized to allow
setting $T_\mathrm{e} = T_\mathrm{f}$. The $\ln{\Lambda}$ factor
accounts for the many weak scatterings out to distance
$\lambda_\mathrm{D}$. The functions $H$ and $\Phi$ are given
by\footnote{In \citep{spitzer56} the function in equation~(\ref{eq:G})
is called $G$, but we rename it $H$ to avoid confusion with the
gravitational constant.}
\begin{eqnarray}
\label{eq:G}
H(\xi) = \frac{\Phi(\xi) - \xi\Phi^{\prime}(\xi)}{2\xi^2}\\
\label{eq:Phi}
\Phi(\xi) = \frac{2}{\sqrt{\pi}}\int_0^\xi e^{-u^2}du.
\end{eqnarray}

Two related questions are considered here. First, for a newly ionized
particle with speed $v_\mathrm{ion}$, how long does it take to
equalize its speed with the rest of the ions? Second, are high-$\beta$
ions dynamically coupled to low-$\beta$ ions?

\begin{figure}
\plotone{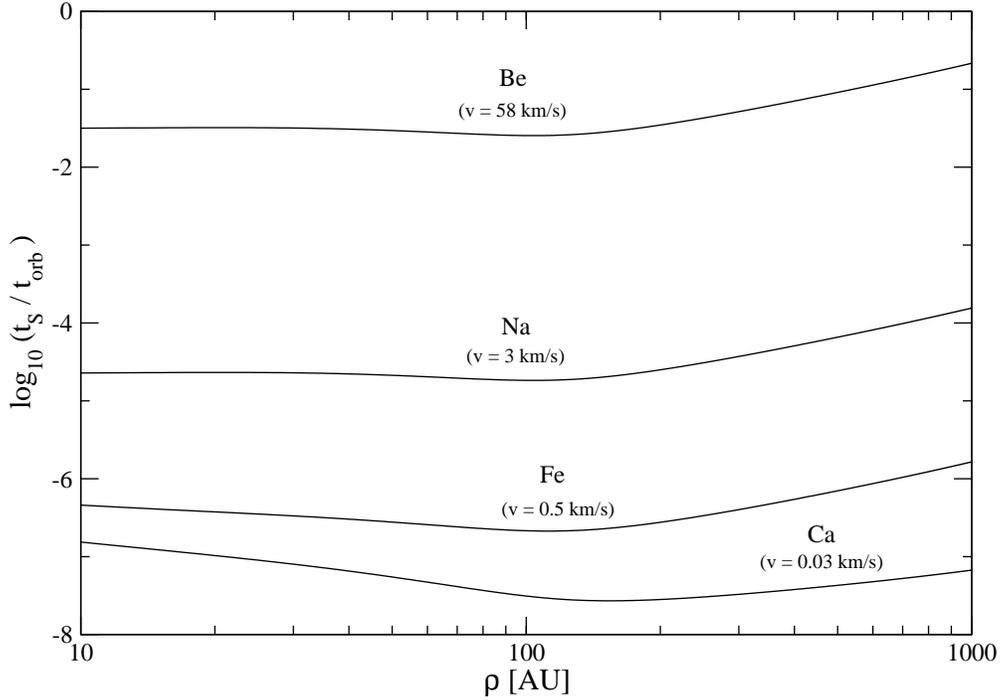}
\caption{Ratio of slow-down to orbital timescales along the disk midplane, for ions with incident velocities $v=v_\mathrm{ion}$,
where $v_\mathrm{ion}$ is the velocity of neutral particles just before ionization, shown in Table~\ref{t:beta_velocities}.
We adopt C\,II as field particles, but the results remain similar if other field particles are used.}
\label{fig:ts_tacc_mid}
\end{figure}

The first question can be answered by evaluating $t_\mathrm{S}$ with
the relative velocity $v=v_\mathrm{ion}$, where $v_\mathrm{ion}$ is
the velocity of neutral particles just before ionization, shown in
Table~\ref{t:beta_velocities}. The resulting $t_\mathrm{S}$ in the
disk midplane is shown in Figure~\ref{fig:ts_tacc_mid} in units of the
local Keplerian orbital period. It takes a very short amount of time
for most newly ionized particles to be equalized with the ionic sea
through Coulomb collisions, except for the element Be. A prediction
is therefore that Be will be preferentially lost from the disk.

\begin{figure}
\plotone{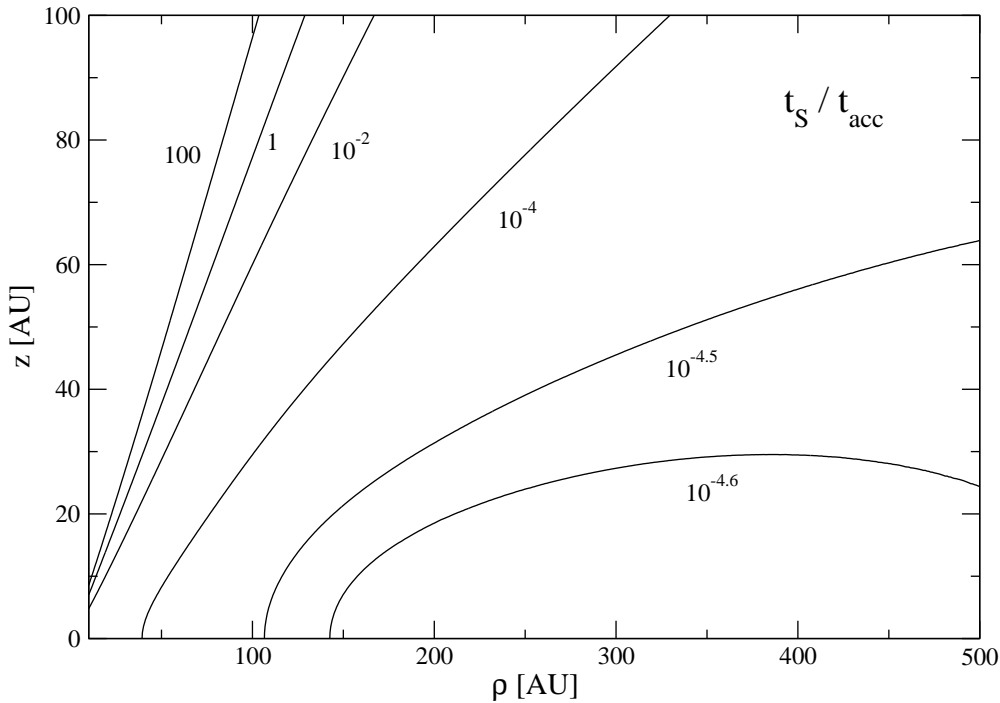}
\caption{Contours of the ratio between the ion slow-down timescale and the radiative acceleration timescale (equations~\ref{eq:t_S} 
and \ref{eq:tacc}) in the $\beta$\,Pic disk ($\rho$ and $z$ are disk
cylindrical coordinates). As an example, we adopt C\,II as field
particles and Fe\,II as test particles with injection velocity $v =
c_\mathrm{th,f}$ ($\xi=1$). In regions of the disk where metallic gas
has been observed, the ion slow-down time is orders of magnitude
shorter than the radiative acceleration time, indicating that ions are
dynamically coupled to a single fluid. Results are similar when we
adopt other elements as field particles.}
\label{fig:ts_tacc}
\end{figure}

To address the second question, we express the timescale to
radiatively accelerate a test ion as
\begin{equation}
\label{eq:tacc}
t_\mathrm{acc} = \frac{m_\mathrm{t} v_\mathrm{*}}{F_\mathrm{rad}} = \frac{r^2 v_\mathrm{*}}{\left(\beta_\mathrm{t} - 1\right) GM},
\end{equation}
where $v_\mathrm{*}$ is the radial velocity relative to the star, and
$\beta_\mathrm{t}>1$ is the test particle radiation force
coefficient. If $t_\mathrm{S}\ll t_\mathrm{acc}$, any excess momentum
acquired by species that feel the radiation force will be transferred
to the field particles as a result of many random collisions, the ions
thus being dynamically coupled despite their different $\beta$
values. Notice that the ratio $t_\mathrm{S}/t_\mathrm{acc}$ does not
depend on $v$ except through the function $H(\xi=v/c_\mathrm{th,f})$,
which reaches a maximum when $\xi\simeq 1$. In
Figure~\ref{fig:ts_tacc}, we plot the ratio
$t_\mathrm{S}/t_\mathrm{D}$ as a function of the disk cylindrical
coordinates $\rho$ and $z$, taking $\xi = 1$, with C\,II as field
particles and Fe\,II as test particles\footnote{When $\xi\ll 1$ or
$\xi\gg 1$, $t_\mathrm{S}$ can become longer that $t_\mathrm{acc}$. In
the former case, ions may drift relative to each other, but with a
speed orders of magnitude below the local sound speed.}. The species
chosen are representative as they are the most abundant (in number and
mass) within groups that feel or do not feel radiation force
($\beta_\mathrm{C\,II}\approx 0$ and $\beta_\mathrm{Fe\,II}\approx 5$,
respectively).

Of particular interest is the coupling of ions at high altitudes over the midplane. \citet{brandeker04} detected Ca\,II emission
being consistent with Keplerian orbits even at a height of $z\sim$80\,AU ($\rho\sim$120\,AU). In that zone we have
$t_\mathrm{S}/t_\mathrm{acc}\approx 0.05$ for $v = c_\mathrm{th,f}$, i.e., ions are still coupled there.
The effect of radiation force on the coupled ensemble of ions can be
quantified by an effective radiation force coefficient
\begin{equation}
\label{eq:beta_eff}
\beta_{\mathrm{eff}} = \frac{\sum_j F_{\mathrm{rad},j}}{\sum_j F_{\mathrm{grav},j}} = \frac{\sum_j \beta_j \varrho_j}{\sum_j \varrho_j},
\end{equation}
where $F_{\mathrm{rad},j}$ and $F_{\mathrm{grav},j}$ are the radiative
and gravitational forces acting on species $j$ within a fluid element,
$\beta_j$ is the radiation pressure coefficient for species $j$, and
$\varrho_j$ is the mass density of species $j$.  Thus
$\beta_{\mathrm{eff}}$ is a weighted average of the values for each
particle, the weight being the mass density of each species. If
$\beta_\mathrm{eff} < 0.5$, ions brake by themselves. Since ions
outweight electrons by a factor $>$2000, and there is one electron per
ion, we neglect the electrons in the estimate of $\beta_\mathrm{eff}$
and assume that they are dragged along the ion fluid, preserving
charge neutrality.

\begin{figure}
\plotone{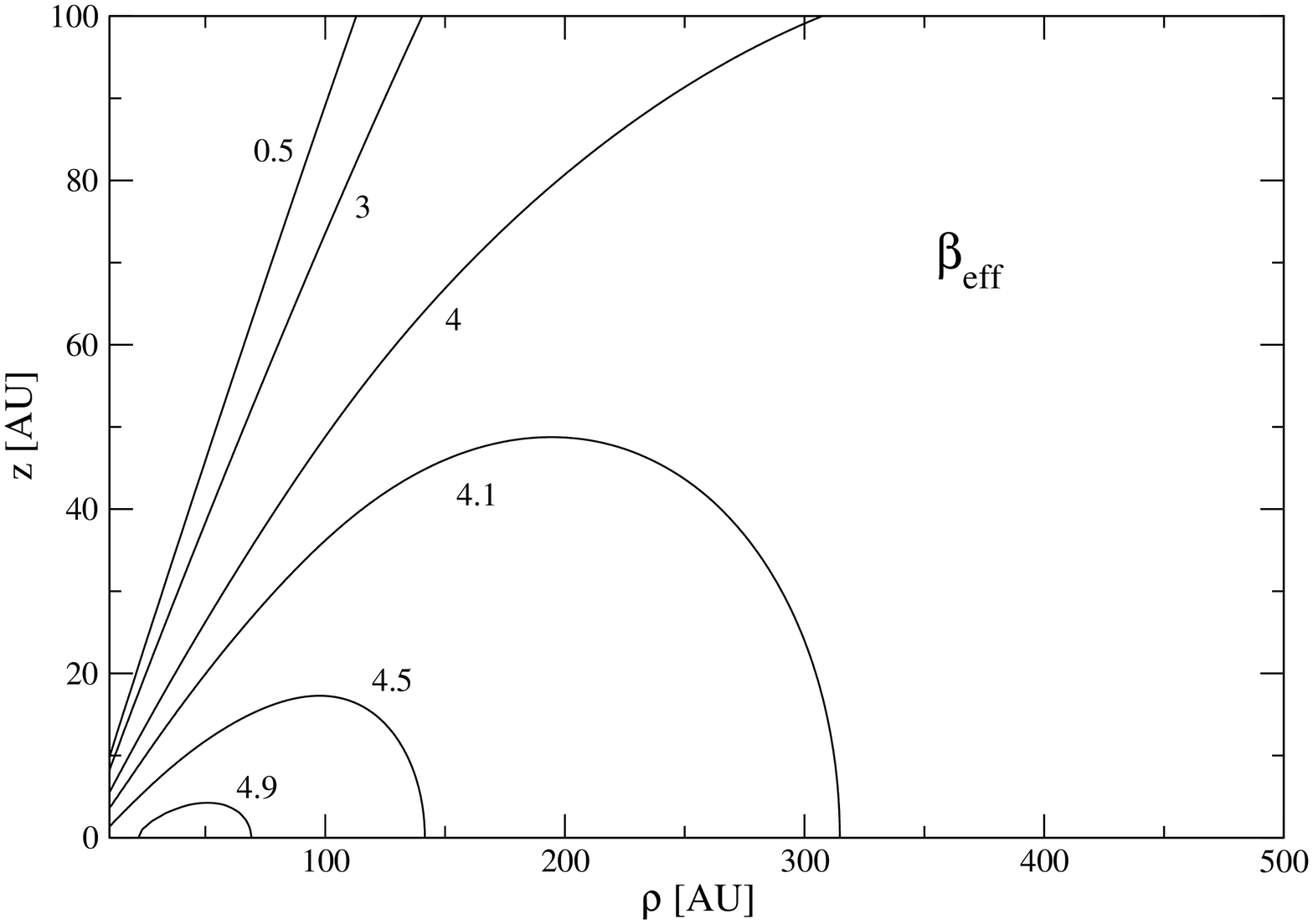}
\caption{Contours of effective radiation pressure coefficient $\beta_\mathrm{eff}$ (equation~\ref{eq:beta_eff}), calculated by 
averaging the radiation force over all ionized particles, as a
function of disk cylindrical coordinates $\rho$ and $z$. Solar
composition is assumed. Carbon, with $\beta=0$, is important for
reducing the value of $\beta_\mathrm{eff}$. Increasing the ionization
as one moves to larger distances (both outward and upward) accounts
for the lowering of $\beta_\mathrm{eff}$. However, $\beta_\mathrm{eff}
> 0.5$ for all regions where metallic gas has been observed,
indicating that an external agent is required to brake a solar-composition 
ionic fluid.}
\label{fig:beta_eff}
\end{figure}

For solar composition, the most abundant ions by mass are, in
decreasing order, C($\beta \approx 0$), Fe(5), Si(9), Mg(9), S(0),
Ni(0) and Ca(50). These particles dominate the value of
$\beta_\mathrm{eff}$, with carbon and iron being the relevant species. 
If all of them were $100\%$ ionized, we would
have $\beta_\mathrm{eff}\approx 3.5$, but since the ionization
fraction of C\,II varies with distance from the star, the actual value
of $\beta_\mathrm{eff}$ is higher (by a factor 2 at most) and has a
weak dependence on position. In Figure~\ref{fig:beta_eff} we plot
$\beta_\mathrm{eff}$ averaged over all ionized species as a function
of disk cylindrical coordinates $\rho$ and $z$. A straightforward way
of braking the gas is by invoking an enhancement in the carbon abundance
by a factor $\sim$10. Observations of circumstellar absorption lines
from C\,I and C\,II by FUSE, reported recently by
\citet{roberge05}, seem to indicate that carbon may indeed be over-abundant 
by this factor. Whether the gas is produced at this abundance or has
evolved to it is, however, not clear at the moment. In the following
sections we proceed to explore additional braking mechanisms that act
independently of the relative abundance pattern of the gas.

\subsection{Collisions with Charged Dust}
\label{s:dustcoll}

Dust grains immersed in a plasma acquire a non-zero charge
\citep{spitzer41}, a phenomenon observed in the interstellar medium
(ISM) and the solar system (e.g., \citealt{mendis94},
\citealt{horanyi96}). We therefore expect grains in the $\beta$\,Pic
system to be charged as well. This might contribute to slow down the
ion fluid by enhancing the ion-grain collision frequency relative to
the case where grains are neutral. The mass in dust grains is expected
to be larger than that of ionized particles, satisfying the
requirement of high inertia. We therefore focus on the requirement of
high collision frequency, which depends on the potential to which
grains are charged to.

\begin{figure*}
\plottwo{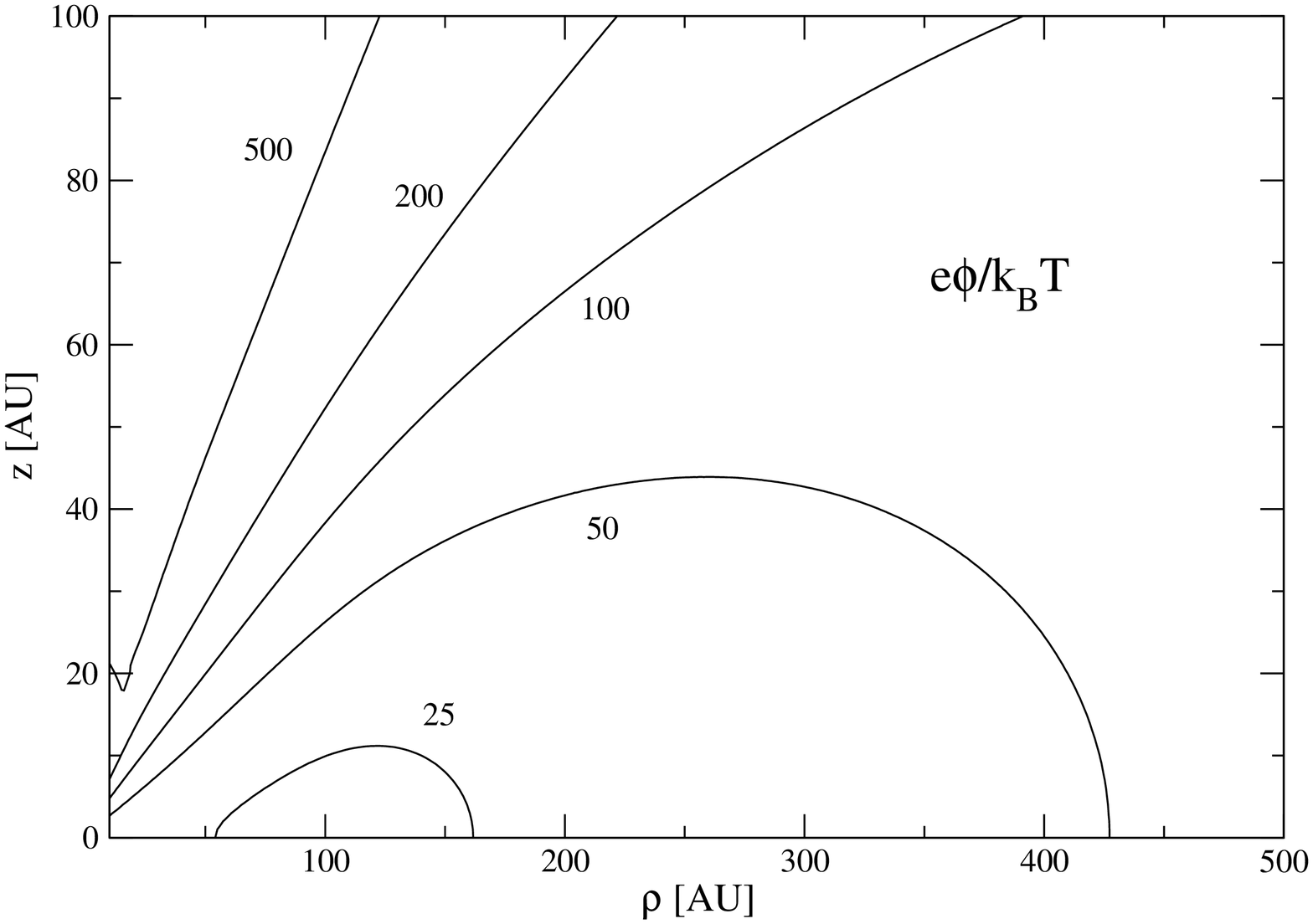}{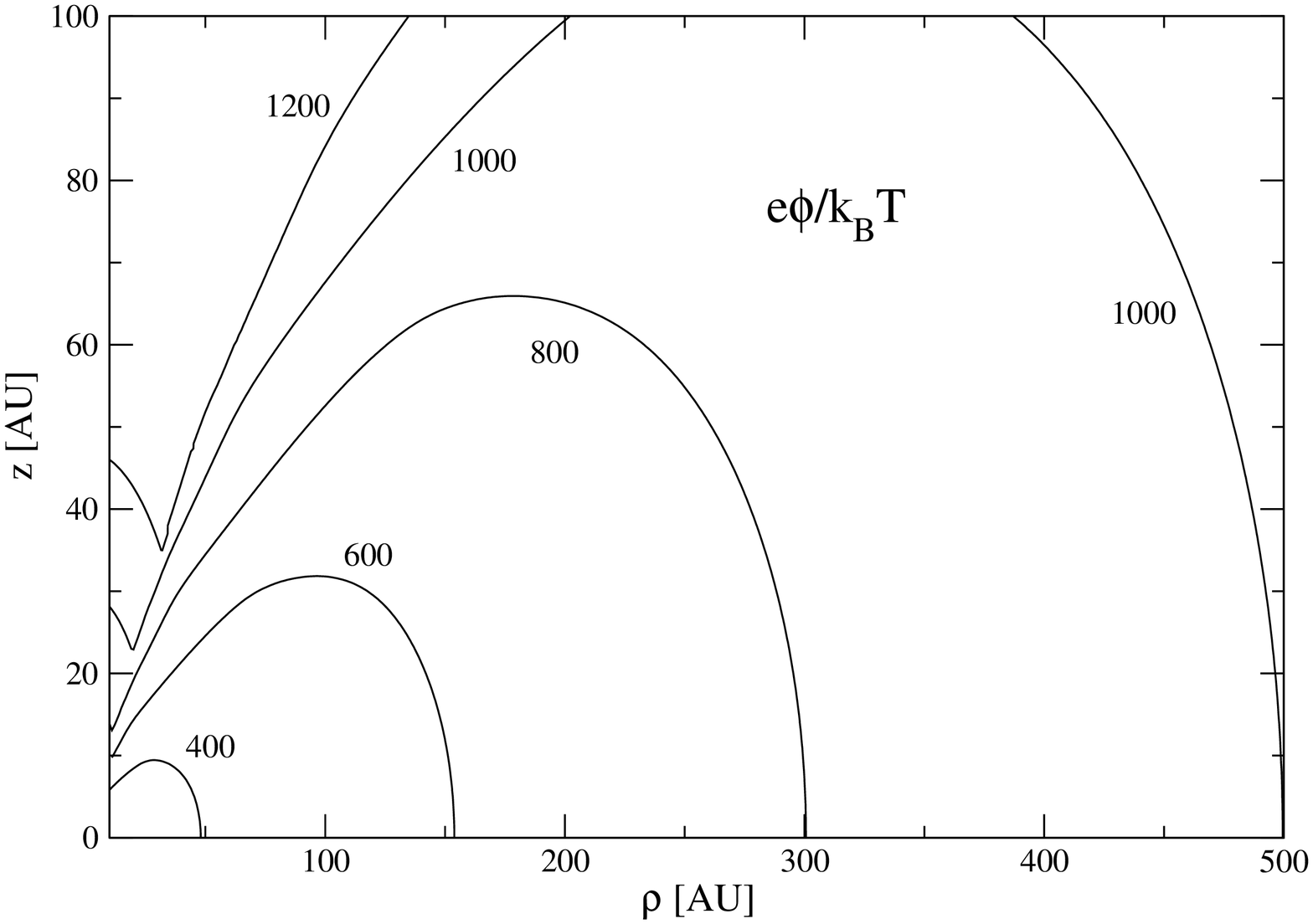}
\caption{Equilibrium electrostatic potential of spherical grains $e\phi$ 
(see Appendix~\ref{s:charge} for details on the calculation) normalized by
$k_\mathrm{B}T$, as a function of disk cylindrical coordinates. Left
panel shows silicates ($W = 8$\,eV) and right panel carbonaceous grains
($W = 4.4$\,eV). Due to strong photoelectric currents, both types of grains are
charged to positive potentials much greater than $\sim k_\mathrm{B}T$, the
value expected if only thermal collection currents are responsible for
charging grains. Results shown here are robust and depend only logarithmically 
on electron density, stellar luminosity, and orbital separation.}
\label{fig:y}
\end{figure*}

The relative importance of different grain charging mechanisms is
determined by environment conditions (e.g., \citealt{mendis94}). In
all cases, an equilibrium electrostatic potential is reached by
balancing positive and negative currents. Given the spectral type of
$\beta$\,Pic, and the fact that the disk is optically thin, the
dominant source of positive charge is the ejection of photoelectrons
by stellar UV radiation, while negative charge is provided by the
collection of thermal electrons. The contribution of thermal ions and
secondary electrons is negligible. As detailed in
Appendix~\ref{s:charge}, we solve for the equilibrium potential of
grains in the $\beta$\,Pic disk given our model spectrum, temperature
profile, and electron density (\S\S
\ref{s:radfor} \& \ref{s:ionstate}). Due to the exponential fall-off of photon flux in the UV
wavelengths, the potential depends weakly (in fact, logarithmically) on
all variables except for the work function $W$ of the grain
(eq. \ref{eq:logarithmic}). More detailed calculations yield that
silicate dust ($W = 8$\,eV,
\citealt{weingartner01}) is charged to a positive potential $e\phi
\sim 0.2$\,eV, while carbonaceous grains ($W = 4.4$\,eV,
\citealt{weingartner01}) achieve a much higher potential $e\phi \sim 2.5$\,eV, 
both values being fairly insensitive to the environment factors. High
value for the latter group results from the fact that stellar photons
with energy $>4.4$\,eV are much more numerous than those above
$8$\,eV. In any case, both potentials are much higher than $\sim
k_\mathrm{B} T \sim 10^{-3}$\,eV, the typical potential achieved when
only thermal collection currents are included. Figure~\ref{fig:y}
shows the potential of silicates and carbonaceous grains normalized by
$k_\mathrm{B} T$, as a function of disk cylindrical coordinates. The
contribution of the ejected photoelectrons to the ambient electron
density is negligible, hence we did not couple grain charging with the
ionization balance calculation.

Each dust grain is orders of magnitude more massive than a single ion,
therefore grains can be considered to be immobile in
space. As shown in \S\ref{s:ioncoll}, ions feel radiation force as an
ensemble. They move with a velocity $v$ relative to the grains, with
an internal velocity dispersion $c_\mathrm{th}$
(equation~\ref{eq:c_thermal}). In a frame in which the center of mass
of ions is at rest, dust grains drift with velocity $-v$.  This
relative motion is dissipated in a dynamical friction timescale
(equation~\ref{eq:t_S}), with dust grains and ions as test and field
particles, respectively. Since the frictional force felt by ions is
equal and opposite to that felt by dust grains, the timescale for
slowing down ions in the stellar frame is given by
\begin{equation}
\label{eq:t_id}
t_\mathrm{id} = \frac{\varrho_\mathrm{ion}}{\varrho_\mathrm{dust}} t_\mathrm{S}(\mathrm{dust,ion})
  = \frac{m_\mathrm{ion}k_\mathrm{B} T v}{8\pi n_\mathrm{dust} Z^2_\mathrm{dust}Z^2_\mathrm{ion}e^4 H(\xi) \ln{\Lambda}},
\end{equation} 
where $\varrho_\mathrm{dust}$ is the volume density of dust grains,
$\varrho_\mathrm{ion}$ is that of ions, and the limit $m_\mathrm{t}\gg
m_\mathrm{f}$ was used in equation~(\ref{eq:t_S}). The grain charge
enters equation~(\ref{eq:t_id}) through $Z_\mathrm{dust}$. The
dynamical friction timescale (equation~\ref{eq:t_S}) takes only into
account the effect of distant encounters, i.e., impact parameters much
larger than the grain size, under the assumption that the cumulative
effect of these collisions outweights that of close encounters
(including impactive ones) in determining the drag force
\citep{spitzer56}. Also, this expression does not include collective
effects among ions (e.g., \citealt{northrop90}). Both these effects
are negligible for the following reasons. First, impact parameters are
of order $b\sim Z_\mathrm{dust}e^2/(k_\mathrm{B}T)$, thus most
collisions are within the range $a\ll b\ll \lambda_\mathrm{D}$, where
$a$ is the grain size and $\lambda_\mathrm{D}$ is the Debye screening
length (consequently, we use $b_\mathrm{min}=a\simeq 1$\,$\mu$m in
equation~\ref{eq:Lambda}). Second, we have
$\omega_\mathrm{p}a/c_\mathrm{th}\sim 10^{-3}$, where
$\omega_\mathrm{p}=(4\pi n_\mathrm{e}e^2/m_\mathrm{e})^{1/2}$ is the
electron plasma frequency, thus collective effects are not likely to
contribute significantly to the drag force.

\begin{figure}
\plotone{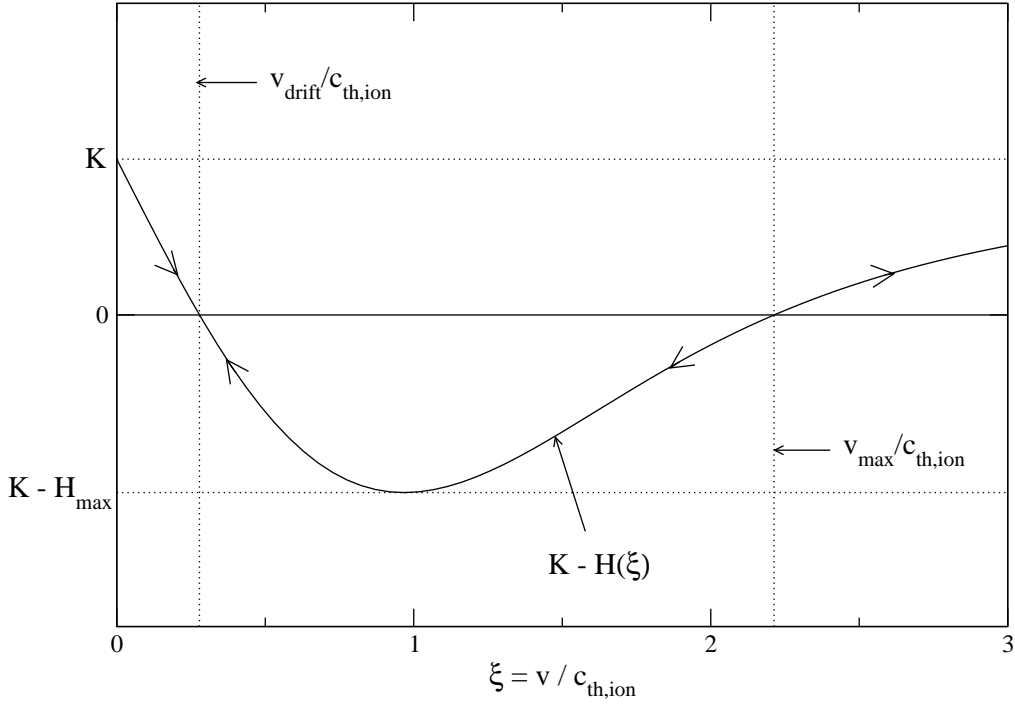}
\caption{Function $K-H(\xi)$ (equation~\ref{eq:motion_dust}), which determines the sign of the net force acting on an ion. Arrows
indicate the stable and unstable nature of the equilibrium velocities
$v_\mathrm{drift}$ and $v_\mathrm{max}$, respectively. For any given
value of $K < H_\mathrm{max}$, $v_\mathrm{max}$ is the maximum
velocity for which braking can occur, whereas $v_\mathrm{drift}$ is
the drift velocity that ions reach in a time
$t_\mathrm{id}$~(equation~\ref{eq:t_id}).}
\label{fig:drag_force}
\end{figure}

Under these assumptions, the equation of motion of a representative
ion is given by
\begin{eqnarray}
m_\mathrm{ion}\frac{dv}{dt} & = & \left( \beta_\mathrm{eff} - 1\right)\frac{GM m_\mathrm{ion}}{r^2} - \frac{m_\mathrm{ion}v}{t_\mathrm{id}}\nonumber\\
\label{eq:motion_dust}
 & = &(\beta_\mathrm{eff}-1)\frac{GMm_\mathrm{ion}}{r^2}\left[\frac{K - H(\xi)}{K}\right],
\end{eqnarray}
where
\begin{equation}
\label{eq:K}
K = \frac{(\beta_\mathrm{eff}-1)GMm_\mathrm{ion} k_\mathrm{B} T}{8\pi
r^2 n_\mathrm{dust} Z^2_\mathrm{dust} Z^2_\mathrm{ion} e^4
\ln{\Lambda}}.
\end{equation}
The function $H(\xi)$ is always positive, with a single maximum
$H_\mathrm{max} \approx 0.2$ at $\xi \approx 1$, and asymptotic limits
$H(\xi)\sim 2\xi/(3\pi^{1/2})$ for $\xi \ll 1$, and $H(\xi)\sim
1/(2\xi^2)$ for $\xi \gg 1$. Therefore, if $K > H_\mathrm{max}$, no
braking is possible, since the net force acting on an ion is always
positive. In this case, as $v$ increases, the relative contribution of
the drag force decreases as $1/v^2$, leading to a runaway solution. If
$K < H_\mathrm{max}$, there exist two equilibrium velocities for which
the net force on the ion vanishes, as shown in
Figure~\ref{fig:drag_force}. We denote them as $v_\mathrm{max}$ and
$v_\mathrm{drift}$ ($v_\mathrm{max} > v_\mathrm{drift}$). Ions with
initial velocities $v_\mathrm{drift} < v < v_\mathrm{max}$ will feel a
drag force until they move with the velocity $v_\mathrm{drift}$
relative to the dust, while those with initial $v>v_\mathrm{max}$ will
be accelerated outward without bound.  These behaviors reflect the
nature of charged dust braking.

We then calculate the value of $K$ for the $\beta$\,Pic dust disk,
taking $m_\mathrm{ion}=m_\mathrm{Fe}$, $Z_\mathrm{ion}=1$, and
$Z_\mathrm{dust} e^2/a = e\phi$. The term $n_\mathrm{dust} a^2$ is
understood to be summed over all grain sizes. As this term also
appears in the dust optical cross section, its value can be obtained
from fits to scattered and thermal emission
\citep{artymowicz89}. In the present work we use an updated 
fitting formula for the dust profile, which assumes a constant grain
albedo 0.5, and is based on HST observations of \citet{heap00}
(P.~Artymowicz, private communication):
\begin{equation}
\pi \langle a^2 \rangle n_\mathrm{dust} = \frac{\tau_0}{W}\left[ \left( \frac{\rho}{\rho_0}\right)^{-4} + \left(\frac{\rho}{\rho_0}\right)^{6}\right]^{-1/2}\exp{\left[-\left( \frac{z}{W}\right)^{0.7}\right]},
\end{equation}
with $\rho_0 = 120$\,AU, $W = 6.6(\rho/\rho_0)^{0.75}$\,AU, and
$\tau_0 = 2\times 10^{-3}$. The dust cross section (or differential
optical depth) peaks at $\sim$120\,AU, with an outward drop-off and an
inner clearing region.

\begin{figure*}
\plottwo{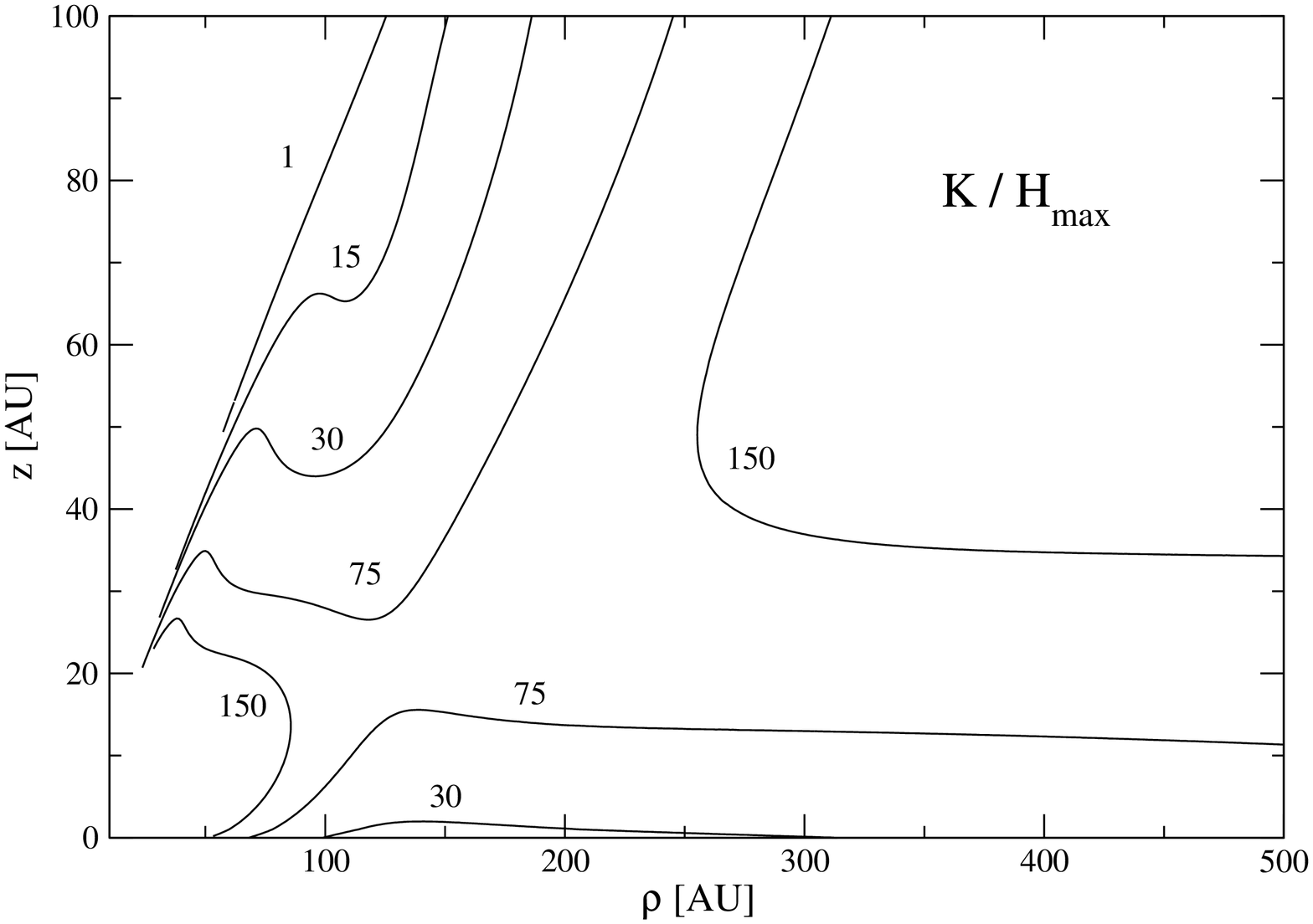}{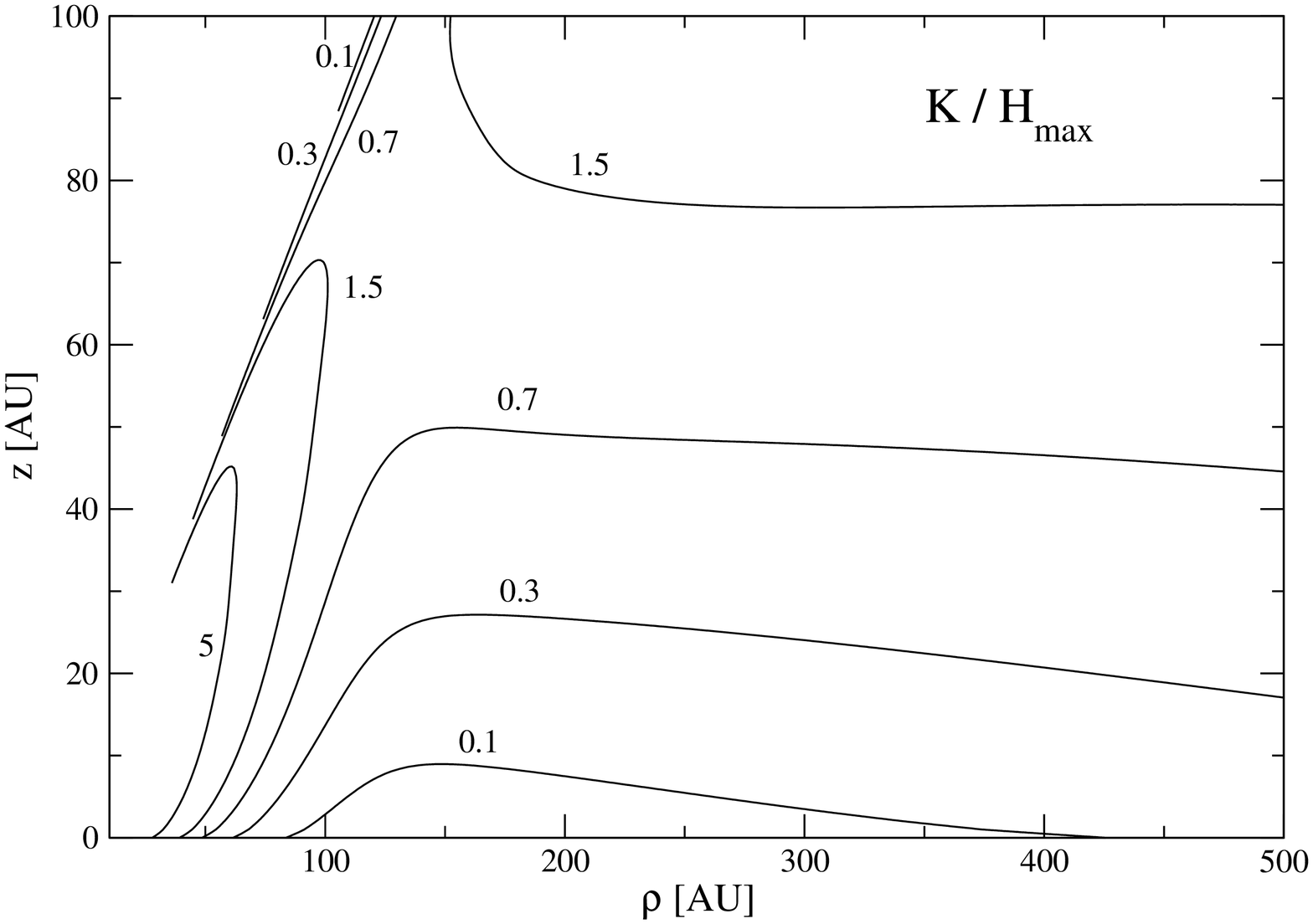}
\caption{Constant $K$ normalized by the maximum value of function $H$ (equation~\ref{eq:motion_dust}) 
for 100\% silicate (left) and 100\% carbonaceous 
grains (right). Braking by charged dust requires $K/H_\mathrm{max} <
1$. While silicate grains are incapable of braking the ions,
carbonaceous grains attain enough charge to exert a significant drag
force and are thus an effective braking agent.}
\label{fig:KGmax}
\end{figure*}

\begin{figure*}
\plottwo{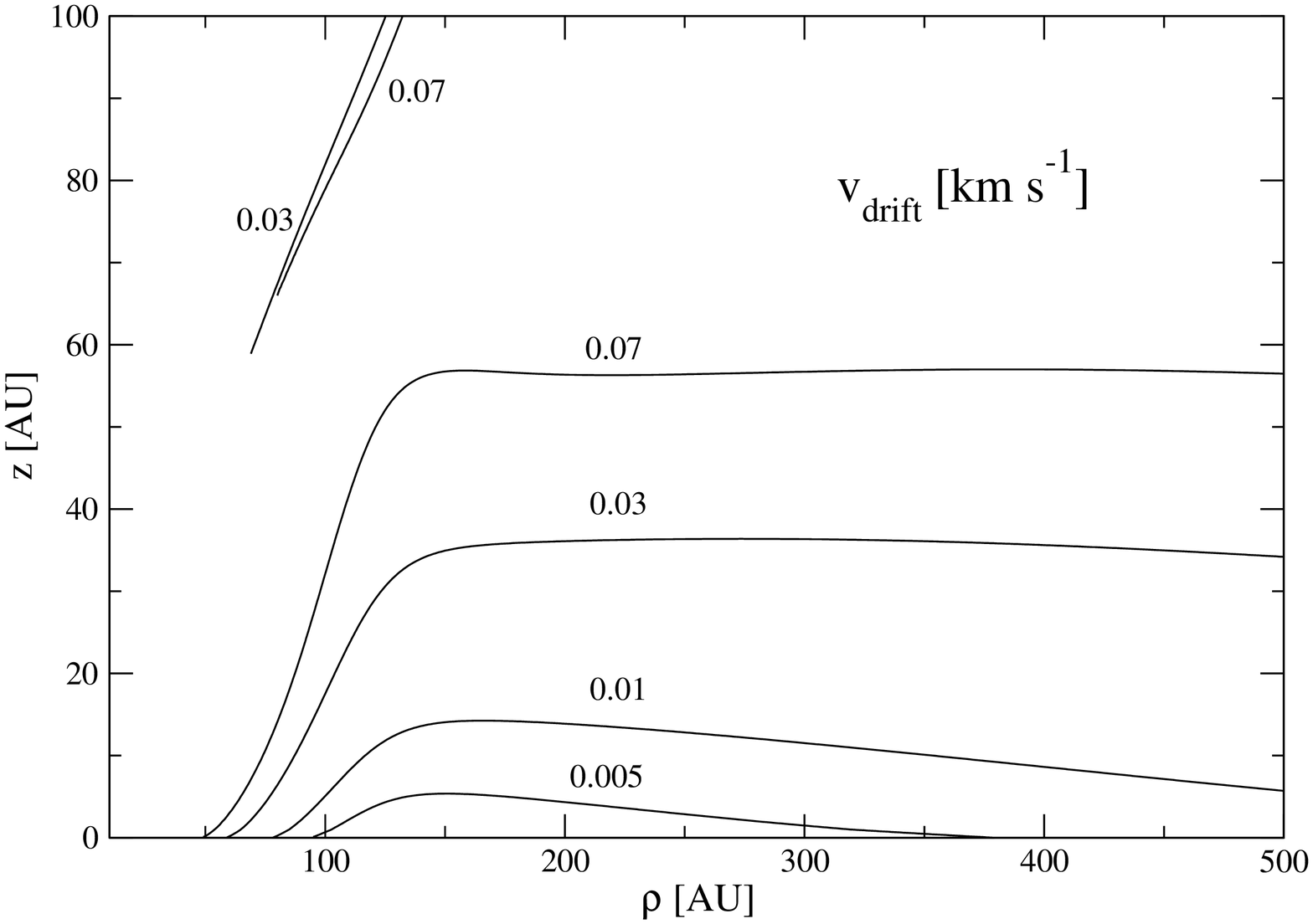}{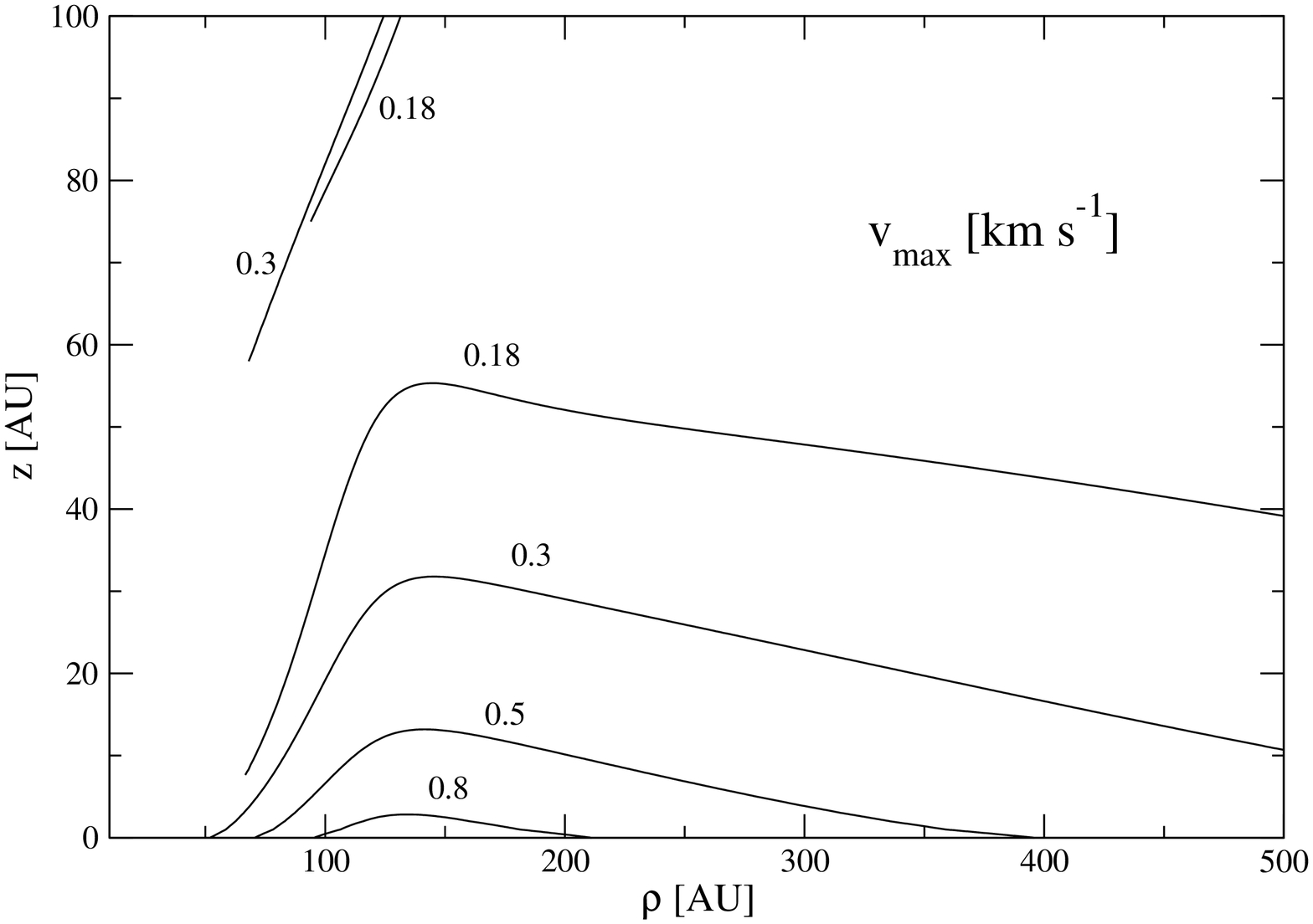}
\caption{Drift velocity $v_\mathrm{drift}$ (left) and maximum velocity for braking $v_\mathrm{max}$ (right) for 100\% 
carbonaceous grains, as a function of disk cylindrical
coordinates. The value of $v_\mathrm{drift}$ falls below observational
limits ($\sim$1\,km\,s$^{-1}$) in the majority of the disk.}
\label{fig:vdrift_vmax_dust}
\end{figure*}

The resulting value of $K$ as a function of disk cylindrical
coordinates is shown in Figure~\ref{fig:KGmax} for 100\% silicate (left)
and 100\% carbonaceous grains (right), normalized by $H_\mathrm{max}$. 
It can be seen immediately
that silicates do not attain enough charge to make their interaction
with ions significant. On the other hand,
carbonaceous grains attain a much greater Coulomb cross section. 
Braking is most effective in the mid-plane region
where dust density is the highest, the efficiency decreasing with
increasing altitude above the midplane. However, at very high
altitudes ($z\geq 80$~AU and $\rho\sim$100~AU, roughly at the location
where Ca\,II emission is detected), somewhat anti-intuitively, dust
braking becomes effective again.  This is explained by
Figure~\ref{fig:y}, which shows increasing grain potential as altitude
rises, a fact that is in turn explained by the exponentially
decreasing electron density with altitude. 
If the disk is made of $100\%$ cabonaceous grains, ions will be slowed
down to velocities compatible with observational constraints, except
very near the star ($\rho\leq 50$\,AU) where the dust density is too
low. Figure~\ref{fig:vdrift_vmax_dust} shows the corresponding values
of $v_\mathrm{drift}$ and $v_\mathrm{max}$ for $100\%$ carbonaceous
grains. 

Are dust grains expected to have a significant fraction of carbon? The
answers seem to depend on the environment in which grains are
found. For ISM grains, \citet{weingartner01} deduced a combination of
carbon and silicates, with carbon taking up $\sim$35\% by volume,
largely what one expects out of solar composition. Chondritic
meteorites have only a few percent of carbon by volume, while most
interplanetary dust particles collected in the Earth stratosphere are
much more carbon rich, with $> 50\%$ of carbon in some cases
\citep{keller94}. Similar enrichment is found in comet Halley. It is
thought that chondrites are relatively poor in volatiles because they
have undergone some processing and are less primitive. Three pieces of
work on $\beta$ Pic may hold some clues to this question: albedo
calculations \citep{artymowicz89} suggest that dust in the outer disk
is likely comprised of bright silicates (or ices) darkened by only a
small amount ($< 1\%$) of carbon materials;\footnote{However, this
result is based on a spatially unresolved SED. Revisiting the problem using
resolved images \citep{heap00} may yield new insights.}
silicate emission features are detected in the inner $\beta$~Pic
disk ($< 20$ AU, \citealt{weinberger}), though absent from the outer part; 
and a super-solar abundance for carbon in the gas phase is indicated by 
a UV absorption study \citep{roberge05}.

What are the constraints one can impose on the relative abundance of
carbonaceous grains if they are solely responsible for braking the
metallic gas? Braking requires $K/H_\mathrm{max} < 1$ where
$K/H_\mathrm{max} \propto 1/n_\mathrm{dust}$. From
Figure~\ref{fig:KGmax}, we have $K/H_\mathrm{max}\leq 0.1$ near the
midplane region ($z\leq$10\,AU) outside $\rho>100$\,AU for the case of
100\% carbonaceous grains. So for this region, a carbon
fraction $\sim 10\%$ is sufficient for braking. This fraction rises with altitude,
inversely proportional to $K/H_\mathrm{max}$.  We recall again that
the dust density profile is still very uncertain, therefore these
numbers should be taken as a very rough estimate.


\subsection{Collisions with Neutral Gas}
\label{s:neutralcoll}

Collisions between neutral and ionized species have previously been
investigated as a braking mechanism in the $\beta$\,Pic system (e.g.,
\citealt{beust89, lagrange98}). The dipole moment that ions induce on
neutrals enhances the collision cross section relative to the
neutral-neutral value (e.g., \citealt{beust89}).  Previous studies,
using neutral-neutral collisions and $\beta$ values for neutral
species, have estimated that $\sim$50\,M$_\earth$ of neutral gas are
required to slow down elements affected by radiation force to within
observational constraints \citep{brandeker04}. Here we perform a new
estimate on the required mass to account for observations, using the
fact that ions are dynamically coupled and are the only species
feeling the radiation force.

Assuming that each ion-neutral collision absorbs $m_\mathrm{n}
v_\mathrm{i}$ momentum from ions, where $m_\mathrm{n}$ is the mass of
neutral particles and $v_\mathrm{i}$ is the velocity of ions, the drag
force exerted on an ion by neutrals is given by (e.g.,
\citealt{beust89})
\begin{equation}
F_\mathrm{drag} = -\varrho_\mathrm{n}\pi \kappa v,
\end{equation}
where, again, the subscripts i and n refer to ions (test) and neutral
(field) particles, respectively, $v$ is the relative velocity between
the species, and $\kappa$ is given by
\begin{equation}
\label{eq:kappa_def}
\kappa = \sqrt{\frac{4 \alpha_\mathrm{n} Z_\mathrm{i} e^2}{m_\mathrm{n}}},
\end{equation}
where $\alpha_\mathrm{n}$ is the polarizability of neutral
particles. Values of $\alpha_\mathrm{n}$ for several neutral atoms and
molecules are shown in Table~\ref{t:polarizabilities}. The equation of
motion for a representative ion is then
\begin{equation}
\label{eq:euler_gasdrag}
m_\mathrm{i}\frac{dv}{dt} = \left(\beta_\mathrm{eff}-1\right)\frac{GMm_\mathrm{i}}{r^2} - \varrho_\mathrm{n}\pi \kappa v.
\end{equation}
This has a single stable equilibrium solution when 
\begin{equation}
\label{eq:v_drift}
v = v_\mathrm{drift} = \frac{(\beta_\mathrm{eff}-1) GMm_\mathrm{i}}{\pi \kappa \varrho_\mathrm{n} r^2}.
\end{equation}
Assuming that during the time $t_\mathrm{drag}$ it takes to slow down
the ion, the distance it travels is much less than the distance to the
star ($v_\mathrm{drift} t_\mathrm{drag}\ll r$), the time dependent
solution to~(\ref{eq:euler_gasdrag}) is given by \citep{liseau03}
\begin{equation}
\label{eq:vdrift_gas_evol}
v(t) \approx v_\mathrm{drift} + \left(v_0 - v_\mathrm{drift}\right)\exp{\left(-t/t_\mathrm{drag}\right)},
\end{equation}
where $v_0$ is the initial relative velocity between the incident ion and the neutral gas, and
\begin{equation}
t_\mathrm{drag} = \frac{m_\mathrm{i}}{\pi\kappa \varrho_\mathrm{n}}
\end{equation}
is the characteristic timescale for momentum damping due to
ion-neutral collisions.

Requiring a fixed drift velocity everywhere in the disk and
integrating the required mass density (from equation~\ref{eq:v_drift})
over a disk with vertical scale height $H$, we obtain a minimum mass
for braking
\begin{eqnarray}
\label{eq:M_min}
M_\mathrm{min} & = & \frac{4GMm_\mathrm{i}}{\kappa
v_\mathrm{drift}}\int_{0}^{H}\int_{\rho_\mathrm{min}}^{\rho_\mathrm{max}}\frac{(\beta_\mathrm{eff}-1)}{r^2}
\rho\, d\rho dz\nonumber\\ & = & 3\times 10^{-2}\textrm{\,M}_\earth
\left( \frac{m_\mathrm{n}}{m_\mathrm{H}}\right)^{1/2}\left(
\frac{\alpha_\mathrm{H\,I}}{\alpha_\mathrm{n}}\right)^{1/2} \left(
\frac{m_\mathrm{i}}{m_\mathrm{Fe}}\right) \left(
\frac{0.1\textrm{\,km\,s}^{-1}}{v_\mathrm{drift}}\right)
\end{eqnarray}
where for numerical evaluation we have taken $H = 100$\,AU,
$\rho_\mathrm{min}=10$\,AU, and $\rho_\mathrm{max} = 1\,000$\,AU. This
result is much lower than previous estimates, for two reasons. First,
since we are braking the ion ensemble, we have
$\beta_\mathrm{eff}\sim$5, as opposed to individual species with
$\beta\sim$300 (Na\,I, Ca\,I), requiring therefore less neutral
material. Second, as only ions need to be braked, the collision
frequency is enhanced relative to the neutral-neutral case due to
induced dipole moment on the neutrals, the increase being of the same
order as the reduction in the $\beta$ value, further reducing the
amount of braking material required to satisfy observational
constraints. However, we note that since equation~(\ref{eq:M_min})
assumes a constant drift velocity everywhere in the disk, the implied
density profile $\varrho_\mathrm{n} \propto 1/r^2$ is
unrealistic. Thus the value in equation~(\ref{eq:M_min}) should be
taken as a lower limit only.
\begin{deluxetable}{lcc}
\tablecaption{Polarizabilities for neutral atoms and molecules\label{t:polarizabilities}}
\tablewidth{0pt}
\tablecolumns{3}
\tablehead{\colhead{Substance} & \colhead{$\alpha_\mathrm{n}$} & \colhead{$\alpha_\mathrm{n}/m_\mathrm{n}$}\\
                               & \colhead{($10^{-24}$ cm$^3$)} & \colhead{($10^{-24}$\,cm$^3$\,mol\,g$^{-1}$)}
}
\startdata
\sidehead{Experimental\tablenotemark{a}}
H$_2$   & $0.787$ & $0.390$\\
H$_2$O  & $1.501$ & $0.083$\\
He      & $0.208$ & $0.052$\\
CO      & $1.953$ & $0.070$\\
\sidehead{Calculated\tablenotemark{b}}
H       & $0.693$ & $0.686$\\ 
C       & $1.755$ & $0.146$\\
N       & $1.046$ & $0.075$\\
O       & $0.678$ & $0.042$\\
\enddata
\tablenotetext{a}{\citet{cccbdb}}
\tablenotetext{b}{Calculated using density functional B3LYP and basis set aug-cc-pVDZ \citep{cccbdb}}
\end{deluxetable}

\begin{figure}
\plotone{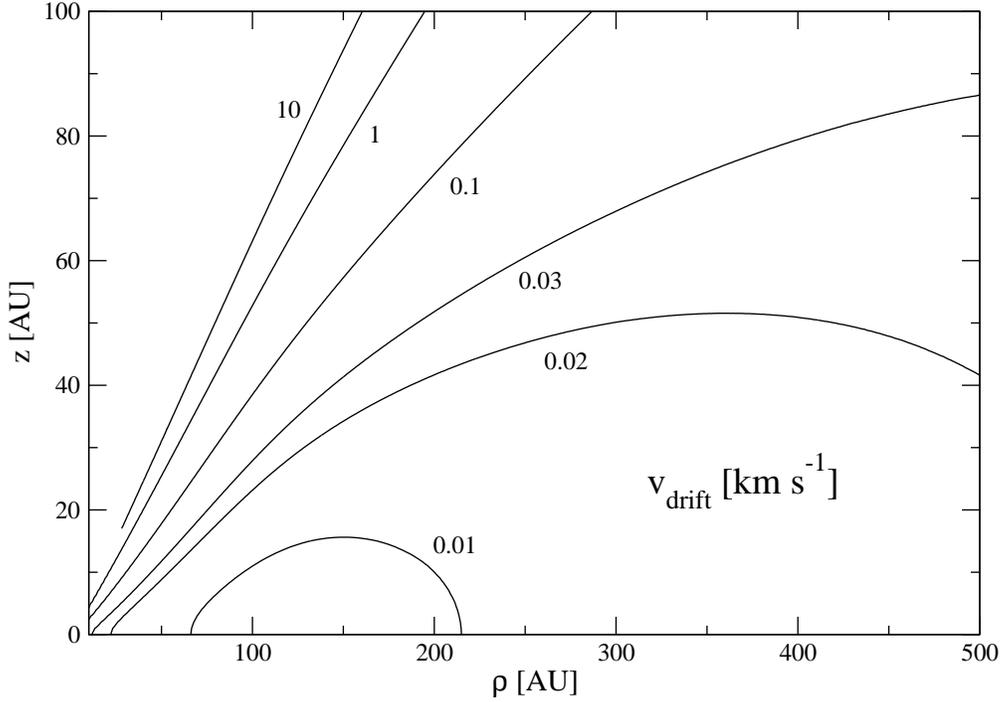}
\caption{Magnitude of equilibrium drift velocity obtained by using the hydrogen density profile of \S\ref{s:ionstate} 
(for solar composition) and equation~(\ref{eq:v_drift}). This
corresponds to a total gas mass of 0.1\,M$_\sun$. While metallic gas
in the midplane region will drift slowly, satisfying observational
constraints, the high-altitude regions where Ca\,II has been detected
($\rho\sim 120$\,AU and $z\sim 80$\,AU) is flowing outwards too fast.}
\label{fig:vdrift}
\end{figure}

Enforcing solar abundances everywhere in the disk, we can calculate
the drift velocity using the density profile of
\S\ref{s:ionstate}, which implies a total disk mass of 0.1\,M$_\sun$. Figure~\ref{fig:vdrift} shows the value of $v_\mathrm{drift}$ 
as a function of disk cylindrical coordinates, assuming Fe\,II as test
particles and H\,I as field particles. In most of the disk, this drift
velocity is well below the observational constraints of
\citet{brandeker04}. However, in the zone corresponding to
high-altitude Ca\,II emission ($\rho\sim 120$\,AU, $z\sim 80$\,AU) we
have $v_\mathrm{drift} \approx 26$\,km\,s$^{-1}$, a factor 100 above
what is observed. This discrepancy has two possible explanations:
either our hydrogen profile falls off too steeply with height (due to,
e.g., a wrong temperature), and/or the solar abundance assumption is
not valid.  The low temperatures in the outer regions of the
$\beta$\,Pic disk allow for the existence of molecules (e.g.,
\citealt{kamp01}).  A possible candidate for a braking medium could
thus be water vapor sputtered from icy dust grains, which certainly
does not need to follow solar abundance relative to the metallic
gas. However, any braking medium other than hydrogen would require a
higher total mass for a given drift velocity, since this element has
the highest $\alpha_\mathrm{n}/m_\mathrm{n}$ ratio and thus highest
$\kappa$, as shown in Table~\ref{t:polarizabilities}. H$_2$O, for
instance, would need to be $\sim$3 times more massive to achieve the
drift velocities in Figure~\ref{fig:vdrift}.

Finally, we address the inertia requirement. Since the drag force on
the ions is equal and opposite to the force exerted by the ions on
the neutral gas, the requirement that the gravitational force on the
latter is much larger than the force from the ions means
$(\beta_\mathrm{eff}-1)\varrho_\mathrm{i}\ll \varrho_\mathrm{n}$, or
$\beta_\mathrm{eff} M_\mathrm{ions}\ll M_\mathrm{min}$. Using the
results of \S\ref{s:ionstate}, we obtain $M_\mathrm{ions}\approx
8\times 10^{-4}$\,M$_\earth$, or $M_\mathrm{min} \approx
40\,M_\mathrm{ions}\sim 10\beta_\mathrm{eff} M_\mathrm{ions}$,
therefore the requirement of high inertia is satisfied even by the
lower limit of equation~(\ref{eq:M_min}). As mentioned before, any
braking medium which follows a physical density profile is likely to
have a mass higher than $M_\mathrm{min}$.

\section{DISCUSSION}
\label{s:discuss}

\subsection{Sensitivity to Gas Temperature}
\label{s:tempsens}

Our above results are obtained by assuming a thermal equilibrium
between gas and dust. How reasonable is this and how sensitive are our
results to this assumption?

\citet{kamp01} made a thorough study of the gas temperature
in the $\beta$\,Pic disk, taking into account relevant heating and
cooling mechanisms. They were able to set an upper limit
$\sim$300\,K for a solar composition gas, the major uncertainty being
the unknown relative drift velocity between dust grains and gas
particles, which they argue to be the dominant heating
mechanism. Our study indicates that the gas disk can be significantly
depleted in hydrogen relative to solar abundance. One naively
expects that the gas will be cooler than the above upper limit, as
cooling is dominated by line emissions from carbon and oxygen. This
needs to be confirmed.

The observed gas disk has a finite thickness, $h/r \sim$0.2 at $100$\,AU 
\citep{brandeker04}.
Assuming that this has a thermal origin (as opposed to a dynamical
origin), and adopting a constant $h/r$ ratio for the whole disk, we
obtain a temperature profile
\begin{equation}
T \sim 9\times 10^4 \textrm{\,K}\, \left({m\over{m_\mathrm{C}}}\right)\,
\left({{h/r}\over{0.2}}\right)^2 \, \left({\mathrm{AU}\over{r}}\right),
\label{eq:Treal}
\end{equation}
where $m$ is the mean mass of a gas particle, and $m_C$ that of a
carbon atom. This yields $T=900$\,K at $r=100$\,AU, much hotter than the
$T=60$\,K value in the $T_\mathrm{gas}=T_\mathrm{dust}$ approximation, and
lies well above the upper limit of \citet{kamp01}. This is puzzling
and deserves further study. 

Even adopting such a seemingly extreme temperature profile, we find
that many of our conclusions remain unchanged. To reproduce the
observed Na\,I density profile, the electron density at 100\,AU rises
slightly from $n_\mathrm{e} = 3.7$\,cm$^{-3}$ to $n_\mathrm{e} =
5$\,cm$^{-3}$. The ionization balance depends on temperature weakly
through the radiative recombination coefficients.\footnote{Given the
low gas densities, collisional ionization, charge exchange
recombinations and dielectronic recombination are irrelevant for the
ionization balance.} As a result, carbon, being intermediately
ionized, is the only element that sees its ionization fraction changed
significantly (rises from $\sim$50\% to $80 - 90\%$). This leads to a
small decrease in the value of $\beta_\mathrm{eff}$ (more C\,II ions
which do not see radiation pressure). The ratio
$t_\mathrm{S}/t_\mathrm{acc}$, relevant for ion coupling, increases by
about an order of magnitude: the linear dependence on $k_\mathrm{B} T$
in the numerator of equation~(\ref{eq:t_S}) is softened by the
increase in the ionized particles (mostly C\,II) and the slow increase
of $\ln{\Lambda}$. Thus, coupling of ions remains efficient. The grain
charge is mildly increased, as electron capture into grains is less
efficient. However, the higher impact velocity increases the ratio
$K/H_\mathrm{max}$ by a factor between $3$ to $5$. Dust braking becomes
less effective, although carbonaceous dust remains capable of braking
gas in the mid-plane. Braking by neutral gas is little affected by
changes in temperature, with a weak indirect dependence via $\beta_\mathrm{eff}$.

\subsection{Gas Origin and Lifetime}
\label{s:origin}

If either ion-grain or ion-neutral collisions are ultimately
responsible for the slowing down of ions in the $\beta$\,Pic system,
there is a finite drift velocity to which particles are slowed
down. An estimate of the lifetime of the metallic gas can be obtained
by calculating the time it takes for a single gas particle to travel
the size of the disk $R$ at a speed $v_\mathrm{drift}$. For charged
dust, assuming a 100\% carbonaceous dust disk, we have
\begin{equation}
t_\mathrm{life}\approx 2\times 10^5 \left( \frac{m_\mathrm{ion}}{m_\mathrm{Fe}}\right)^{1/2}
\left(\frac{60\textrm{\,K}}{T} \right)^{1/2} \left( \frac{0.1}{K/H_\mathrm{max}} \right) \left( \frac{R}{300\textrm{\,AU}} \right)\textrm{\,yr},
\end{equation}
whereas for ion-neutral collisions we have
\begin{equation}
\label{eq:tlife_drag}
t_\mathrm{life}\approx 1.4\times 10^4 \left( \frac{m_\mathrm{H}}{m_\mathrm{f}}\right)^{1/2}
\left(\frac{\alpha_\mathrm{f}}{\alpha_\mathrm{H}} \right)^{1/2} \left( \frac{m_\mathrm{Fe}}{m_\mathrm{t}} \right) 
\left( \frac{R}{300\textrm{\,AU}} \right) \left( \frac{M_\mathrm{min}}{3\times10^{-2}\textrm{\,M}_\earth} \right)\textrm{\,yr}.
\end{equation}
One key question, important for constraints on gas evolution, is where the observed gas originates. Clues may be found in the
observational evidence:
\begin{itemize}
\item[(a)] The metals observed in absorption follow \emph{approximately} solar abundance \citep{lagrange95}.

\item[(b)] There is a strong upper limit on the H$_2$ column density in the disk from FUSE observations 
\citep{lecavelier01} that corresponds to a disk mass $\sim$0.1\,M$_\earth$.

\item[(c)] The metallic gas spatial distribution is very similar to that of the dust \citep{olofsson01,brandeker04}. 

\item[(d)] The metals observed in absorption and emission are close to rest relative to the star \citep{lagrange98,brandeker04}.

\end{itemize}

As to the origin of the observed metallic gas, we see three different possibilities:
\begin{enumerate}
\item The gas could be primordial, that is, a remnant from the initial cloud out of which the star formed. Equation~(\ref{eq:tlife_drag})
then implies a gas mass of 30\,M$_\earth$ to prevent the ions from
leaving the system on a timescale shorter than the lifetime of the
system, $t_\mathrm{age}\approx 10^{7}$\,yr \citep{zuckerman01}. If the
disk gas consists predominantly of hydrogen, then it would be in
disagreement with (b). There remains the possibility that the gas disk
is dominated by elements other than H, such as oxygen or
oxygen-bearing molecules, which could have avoided detection.  From an
evolutionary point of view, however, it may be hard to justify why
there should be such a large reservoir of oxygen left, and little
hydrogen. Furthermore, the recent dynamical modeling of dust behavior
in a gaseous disk by \citet{thebault05} excludes a gas disk more
massive than $\sim$0.4\,M$_\earth$.

\item The gas could be produced by bodies evaporating as they fall into the star, and then blown/diffused out by the radiation
pressure to large distances in the disk
\citep{lagrange98}. \emph{Braking} gas could not be produced in such a
scenario, since the gas would then just accumulate close to the
release radius near the star. The hypothesis of remnant primordial gas
braking the metallic ions from the inner disk faces a fine-tuning
problem: with too little braking gas the metals will leave too
quickly, whereas with too much braking gas the metals would never have
reached the outer disk.

The production rate of gas required, $\dot{M} =
M_\mathrm{ion}/t_\mathrm{life}\sim 10^{-8} -
10^{-9}$\,M$_\earth$\,yr$^{-1}$, is consistent with the production
inferred from the FEB scenario \citep{lagrange88}. According to
\citet{thebault03}, this high evaporation rate is inconsistent with
the structure of the inner dust disk around $\beta$\,Pic. A lower
evaporation rate could still be consistent with the observed metallic
gas density profile, provided that the drift velocity is decreased by
the same amount by increasing the braking gas density. Constraint (b)
limits the braking gas increase to a factor $\lesssim 3$.

If charged dust is primarily responsible for the gas braking, then (c)
is a natural consequence of this. However, braking by dust is unlikely
in this case. The main reason is that the gas is initially injected
with a velocity $\gg v_\mathrm{max}$
(Figure~\ref{fig:vdrift_vmax_dust}), even in the favorable case of
carbonaceous grains.

\item The gas could be created in grain-grain collisions, in line with the production of small dust grains from minor bodies
\citep{thebault03}. The correlation between the dust and gas spatial distribution (c) would be a natural consequence of the 
production mechanism. Braking by dust is possible but not required,
since even if the metallic ions are rapidly depleted initially, the
neutral low-$\beta$ gas is left behind until the appropriate abundance
pattern is reached for the ions to self-brake. If this is the case,
then the gas volatiles are predicted to be somewhat enhanced relative
to the metals, and at the same time deficient in H, since grains are
expected to contain very little hydrogen.

If grains are able to brake ions, not only would the ions be tracing
out the space distribution of smaller grains, they should also reveal
the kinematics of dust. The near-Keplerian profile of metallic ions in
the $\beta$\,Pic disk would therefore indicate that the grains, even
at 300\,AU, are still orbiting the star, raising the issue of how
these grains get to these remote locations, since radiative blow out
is not an option.
\end{enumerate}

In summary, our results indicate that the hypothesis that the observed
metallic gas is primordial can be excluded. We cannot rule out the
possibility that the metallic gas is produced by falling evaporating
bodies and then diffused outwards, but this requires coincidental
circumstances that need a priori justification. We conclude that the
gas is most likely produced by grain evaporation (possibly due to
grain-grain collisions) and predict a gas composition where metals are
depleted relative to the volatiles, and where the volatiles have the
same relative abundance as in the grains.

\subsection{Extension to other systems}
\label{s:extension}

Given that several debris disks have been discovered around main
sequence stars with different spectral types, we may ask whether the
braking mechanisms we have studied so far are also important in those
systems.  We can do the simple exercise of calculating the radiation
force coefficients, the ionization state, the value of
$\beta_\mathrm{eff}$, and attempt to draw some qualitative conclusions
based on what we have learned in the previous
sections. Table~\ref{t:othersystems1} shows radiation force
coefficients and neutral fractions at $r=100$\,AU in the midplane, as
a function of spectral type, for C, Fe, Si, Mg, S, Ni, Ca, Al, Na, and
K. The spectra are high resolution PHOENIX atmosphere models
(P. Hauschildt, private communication) similar to those described in
\S\ref{s:radfor}, flux calibrated and rotationally broadened according
to the following stars: HR\,4796A~(A0V, 3\,$M_\sun$, 55\,L$_\sun$,
$v\sin{i} = 152$\,km\,s$^{-1}$), HD\,107146~(G2V, 1\,M$_\sun$,
1\,L$_\sun$, 200\,km\,s$^{-1}$), TW\,Hya~(K8V, 0.6\,M$_\sun$,
0.1\,L$_\sun$, 5\,km\,s$^{-1}$), and AU\,Mic~(M1V, 0.5\,M$_\sun$,
0.05\,L$_\sun$, 7\,km\,s$^{-1}$). The ionization state for each case
was obtained under the same constraints as the $\beta$\,Pic data (A5V,
1.75\,M$_\sun$, 11\,L$_\sun$, 130\,km\,s$^{-1}$), which is shown for
comparison.

\begin{deluxetable*}{lccccccccccc}
\tablecaption{Radiation force, neutral fractions\tablenotemark{a}, and velocities before ionization for different
spectral types\label{t:othersystems1}}
\tablewidth{0pt}
\tablehead{\colhead{Type\tablenotemark{b}} & \colhead{Quantity} & \colhead{C} & \colhead{Fe} & \colhead{Si} & \colhead{Mg} & \colhead{S} & \colhead{Ni}
& \colhead{Ca} & \colhead{Al} & \colhead{Na} & \colhead{K}}
\startdata
A0V & $\beta_\mathrm{I}$             & 3       & 60      & 17      & 200     & 12      &  48      &   580   &  99     &   420   & 200\\ 
    & $\beta_\mathrm{II}$            & 0.1     & 21      & 25      & 40      & 0.01    &  0.5     &   180   &  20     &   0     & 0\\ 
    & $n_\mathrm{I}/n_\mathrm{II+}$  & 0.01    & 3E-6    & 2E-7    & 6E-6    & 5E-4    &  5E-6    &   5E-7  &  1E-7   &   4E-5  & 3E-5\\
    & $v_\mathrm{ion}$ [km\,s$^{-1}$] & 2       & 5E-3    & 1E-4    & 0.04    & 0.02    &  6E-3    &   0.01  & 2E-4    &   0.06  & 0.1\\
A5V & $\beta_\mathrm{I}$             & 0.03    & 27      & 6       & 74      & 0.6     &  26      &   330   &  53     &   360   & 200\\ 
    & $\beta_\mathrm{II}$            & 2E-3    & 5       & 9       & 9       & 9E-5    &  0.07    &   53    &  0.4    &   0     & 0\\ 
    & $n_\mathrm{I}/n_\mathrm{II+}$  & 0.6     & 9E-4    & 1E-4    & 6E-4    & 0.06    &  5E-4    &   4E-6  &  9E-7   &   3E-4  & 2E-4\\ 
    & $v_\mathrm{ion}$ [km\,s$^{-1}$] & \nodata & 0.5     & 0.02    & 1       & 2       &  0.3     &   0.03  & 5E-4    &   3     & 0.5\\
G2V & $\beta_\mathrm{I}$             & 1E-5    & 0.06    & 4E-3    & 0.2     & 2E-5    &  0.1     &   3     &  0.3    &   12    & 26\\ 
    & $\beta_\mathrm{II}$            & 1E-7    & 7E-3    & 0.01    & 0.06    & 3E-8    &  2E-6    &   0.8   &  6E-5   &   0     & 0\\ 
    & $n_\mathrm{I}/n_\mathrm{II+}$  & 0.6     & 0.9     & 0.06    & 2       & 0.2     &  2       &   6E-3  &  1E-3   &   0.08  & 0.03\\ 
    & $v_\mathrm{ion}$ [km\,s$^{-1}$] & \nodata & \nodata & \nodata & \nodata & \nodata &  \nodata &   0.2   & \nodata &   17    & 7\\
K8V & $\beta_\mathrm{I}$             & 6E-9    & 4E-3    & 4E-5    & 2E-3    & 3E-8    &  7E-3    &   0.2   &  0.02   &   3     & 11\\ 
    & $\beta_\mathrm{II}$            & 2E-11   & 2E-4    & 4E-5    & 2E-3    & 2E-     &  6E-8    &   0.06  &  3E-8   &   0     & 0\\ 
    & $n_\mathrm{I}/n_\mathrm{II+}$  & 0.7     & 1       & 0.07    & 3       & 0.2     &  3       &   0.5   &  0.1    &   14    & 13\\ 
    & $v_\mathrm{ion}$ [km\,s$^{-1}$] & \nodata & \nodata & \nodata & \nodata & \nodata &  \nodata & \nodata & \nodata &   9E+3  & 2E+3\\
M1V & $\beta_\mathrm{I}$             & 7E-10   & 5E-4    & 6E-6    & 4E-4    & 8E-9    &  7E-4    &   0.04  &  2E-3   &   0.6   & 2.1\\ 
    & $\beta_\mathrm{II}$            & 4E-12   & 2E-5    & 7E-6    & 2E-4    & 6E-10   &  2E-8    &   0.01  &  3E-9   &   0     & 0\\ 
    & $n_\mathrm{I}/n_\mathrm{II+}$  & 0.7     & 1       & 0.07    & 3       &   0.2   &  3       &   0.6   &  0.1    &   15    & 20\\ 
    & $v_\mathrm{ion}$ [km\,s$^{-1}$] & \nodata & \nodata & \nodata & \nodata & \nodata & \nodata  & \nodata & \nodata &  1E+4   & 2E+3\\
\enddata
\tablenotetext{a}{Neutral fractions were computed at $r=100$\,AU in the disk midplane, under the same constraints as the 
values determined in \S\ref{s:ionstate} for $\beta$\,Pic, changing only the spectrum.}
\tablenotetext{b}{Corresponding to model spectra (see text for details), flux calibrated and rotationally broadened according to
particular cases (see text). No chromospheric emission was included in
the model.}
\end{deluxetable*}

As expected, while $\beta$ values decrease with later spectral type,
neutral fractions increase. Spectral types G2, K8 and M1 have
radiation force coefficients much smaller than 1, except for Na\,I and
K\,I. These species are expected to be depleted in stars around K8,
since they are mostly neutral, while still having $\beta>0.5$. Late
spectral types also have emission lines arising from chromospheric
activity, which we do not account for in these calculations. This
might increase the $\beta$ values for the M1V type, making Na and K
depletion more likely in these systems.

For the A0V case, all species are mostly in the ionized phase, but
with higher $\beta$ values than in $\beta$\,Pic. Using the densities
of ionized species obtained by changing the spectral type to A0V, one
obtains $\beta_\mathrm{eff}\approx 15$ at $r=100$\,AU.  This result
should be taken only as a rough reference, since ionization fractions
depend on the densities of different species, which in turn depend on
disk mass and composition.

Also shown are velocities before ionization $v_\mathrm{ion}$
calculated as in equation~(\ref{eq:v_ion}), which have a physical
meaning only when $\beta > 0.5$ (when $\beta <0.5$, the species feels
a gravitational attraction weaker by a factor $[1-\beta]$, but is
still bound to the star). Again, Na and K seem to be the problem,
being likely decoupled from the rest of the ions in the G2 disk, in
the same way as Be is decoupled in $\beta$\,Pic
(\S\ref{s:ionstate}). Ions are likely to be very strongly coupled for
the A0 case.

\section{CONCLUSIONS}
\label{s:conclusions}

Motivated by the apparent contradiction between the strong radiation
force acting on the gas in the $\beta$~Pictoris disk and spatially
resolved observations showing it to be consistent with Keplerian
rotation, we have explored different braking mechanisms that are
likely to operate in the disk. We have found that:
\begin{itemize}
\item[a)] All species affected strongly by radiation force are heavily ionized. The velocities that the short-lived neutral 
particles achieve as a result of radiative acceleration, before
becoming ionized, are consistent with observational constraints.
\item[b)] Ions are dynamically coupled due to their high Coulomb collision frequency. They feel a single effective radiation
force coefficient $\beta_\mathrm{eff}$. For solar composition, this
coefficient is $\sim$5, in which case ions cannot brake by
themselves. If carbon is over-abundant by a factor $\gtrsim$10 on the
other hand, the ion fluid may indeed be self-braking.
\item[c)] If a significant fraction of the dust in the disk is carbonaceous, the ion ensemble can be slowed down to drift
velocities below the local sound speed by photoelectrically charged
dust grains. 
This fraction is as low as $10\%$ near the midplane and
rises with altitude above the midplane.
\item[d)] Ions can also be slowed down by collisions with neutral gas (e.g., H$_2$, H$_2$O). 
The amount of neutral material required to satisfy observational
constraints on drift velocities is $<$1\,M$_\earth$, substantially
less than previous estimates.
The UV absorption upper limit on $H_2$ is $\sim 0.1 M_\earth$ \citep{lecavelier01}.
\item[e)] Reasonable assumptions on gas drift velocities imply gas lifetimes of order $10^4-10^5$\,yr, much lower than 
the age of the system and which require a replenishment mechanism, in
analogy with the dust.
\end{itemize} 

The required gas production rate $10^{-13}$\,M$_\sun$\,yr$^{-1}$ is
comparable to the FEB scenario prediction (e.g.,
\citealt{lagrange88}), although we argue that gas is more likely to be
produced by local
dust grain evaporation. A primordial origin of the gas is ruled out,
since for having a lifetime of $\sim$10$^7$\,yr it would require a
substantial quantity of neutral material, which would have already
been detected.

Further gas observations in this and other systems would be able to
test several predictions arising from this work, such as the absence
of P and Be in $\beta$\,Pic, and the relative importance of ion-grain
and ion-neutral collisions for slowing down ions. A more detailed
study of the dust composition would shed light on the feasibility of
the dust braking scenario.

\acknowledgments

We are grateful to Peter~Hauschildt and Inga~Kamp for making the
high-resolution model spectra available to us. We also thank useful
discussions with Norm~Murray, Pawel~Artymowicz, Chris~Matzner,
Peter~Goldreich, Yuri~Levin, Ren\'e~Liseau, Per~Carlqvist,
Hilding~Neilson, and Amr~El-Zant. Comments by an anonymous referee
were helpful for improving this paper. We acknowledge financial 
support from NSERC. This work made extensive use of
NASA's Astrophysics Data System.

\appendix

\section{UNCERTAINTIES IN THE RADIATION FORCE COEFFICIENTS}
\label{s:uncertainty}

The errors in the radiation force coefficients being quoted in Table~\ref{t:betavalues} were calculated using the following
expression:
\begin{equation}
\left(\frac{\delta \beta}{\beta}\right)^2 = \left(\frac{\delta F}{F}\right)_\mathrm{cal}^2
+ \frac{1}{\beta^2}\sum_i \left( \frac{g_0}{\sum g_0}\right)^2 \beta_i^2 \left[ \left(\frac{\delta A_{i0}}{A_{i0}} \right)^2 + 
\left(\frac{\delta F}{F}\right)^2_{\mathrm{limb},i} \right]
\end{equation}
where $\delta \beta$ is the error quoted in Table~\ref{t:betavalues},
$\beta_i$ and $\delta A_\mathrm{i0}$ are the radiation force
coefficient and error in Einstein coefficient corresponding to the
$i$-th transition, respectively, $(\delta F/F)_{\mathrm{limb},i}$ is
the fractional error in the stellar flux (at the given transition) due
to uncertain limb darkening used in the rotational broadening of the
spectrum, and $(\delta F/F)_\mathrm{cal}$ is the fractional
uncertainty in the overall flux calibration. The term $(g_0/\sum g_0)$
represents the weighted average between different multiplets of ground
states, when applicable (e.g., Fe\,I), where it is assumed that the
population of a multiplet is proportional to its statistical weight.

Values for $(\delta A/A)$ range from 3\% for the stronger transitions,
to more than 50\% for the weaker ones \citep{asd}. The factor $(\delta
F/F)_{\mathrm{limb},i}$ was estimated by rotating the atmosphere model
with limb darkening coefficients $\varepsilon=\{0,0.5,1\}$ and
calculating the fractional deviation, as a function of wavelength,
relative to the $\varepsilon=0.5$ case. As mentioned in the main text,
the error $(\delta F/F)_\mathrm{cal}$ was estimated to be 4\%,
although systematic deviations may arise between the generic model
spectrum and the particular $\beta$\,Pic case.

\section{GRAIN CHARGING IN THE $\beta$~PIC DISK}
\label{s:charge}

The relative importance of different grain charging mechanisms is
determined by four parameters: the dust grain size $a$, the Debye
screening length $\lambda_\mathrm{D}$ of electrons, the mean
intergrain separation $n_\mathrm{d}^{-1/3}$, and the UV flux incident
on the system \citep{mendis94}. A lower limit on the intergrain
separation can be estimated by using the smallest grain size and the
highest dust mass in the disk. Assuming a higher limit for dust mass
$\sim$120\,M$_{\earth}$ in the disk \citep{artymowicz97}, using the
smallest grain size $a\sim$1\,$\mu$m \citep{backman93}, grain mass
density $\sim$1\,g\,cm$^{-3}$, and a disk volume
$\pi(100\mathrm{\,AU})^2\times(10\mathrm{\,AU})$ we estimate
$n_\mathrm{d}^{-1/3}\gtrsim 180$\,cm. Using the results of
\S\ref{s:ionstate}, we get $\lambda_\mathrm{D} \approx
10$--$80$\,cm. Thus, the $\beta$\,Pic disk is in the regime $a <
\lambda_{\mathrm{D}} < n_\mathrm{d}^{-1/3}$, where grains can be
treated as isolated particles immersed in an ionized gas, i.e.,
collective effects of charged grains play no role \citep{mendis94}.

The equilibrium electrostatic potential of a grain $\phi$ is found by
balancing all the currents that charge the grain. Given the UV flux of
the star, photoelectric charging is the dominant process, with a
current per unit area given by (e.g, \citealt{draine78})
\begin{equation}
\label{eq:J_ph}
J_\mathrm{ph} = \frac{e}{4}\int^{\nu_\mathrm{max}}_{(e\phi + W)/h}\left[\int^{h\nu-W}_{e\phi} f(E,h\nu)\, dE\right]Q_\mathrm{abs}(\nu) Y(h\nu) \frac{F_\mathrm{\nu}}{h\nu} d\nu
\end{equation}
for the case of spherical grains with $\phi>0$. Here, $F_\nu$ is the
stellar flux per unit frequency, $Q_\mathrm{abs}$ the grain absorption
efficiency, $Y$ the photoelectric yield, $W$ the grain work function,
and
\begin{equation}
f(E,h\nu) = \frac{6}{h\nu - W}\left(\frac{E}{h\nu - W}\right)\left(1 - \frac{E}{h\nu - W}\right)
\end{equation}
is the photoelectron kinetic energy distribution function, which we
approximate as a parabolic function (e.g.,
\citealt{weingartner01}). The integral over electron energy is then
\begin{equation}
\int_{e\phi}^{h\nu-W}f(E,h\nu)\,dE = 1 - 3\left( \frac{e\phi}{h\nu - W}\right)^2 + 2\left( \frac{e\phi}{h\nu - W}\right)^3.
\end{equation}
The photoelectric yield is approximated as \citep{draine78}
\begin{equation}
Y(h\nu) = \frac{1}{2}\left(1 - \frac{W}{h\nu}\right).
\end{equation}
The fact that the grains in the disk have a minimum size of order
$\sim$1\,$\mu$m \citep{backman93} allows the use of work functions
corresponding to bulk matter. The values adopted are $W = 8$\,eV for
silicates and $W = 4.4$ for carbonaceous grains
\citep{weingartner01}. The minimum size of the grains also justifies taking the absorption efficiency for UV radiation as 
as $Q_\mathrm{abs}(\nu)=1$ (e.g., \citealt{greenberg71}).

The positive photoelectric current in equation~(\ref{eq:J_ph}) is balanced by a thermal electron collection 
current per unit area (e.g., \citealt{spitzer41}):
\begin{equation}
\label{eq:J_e_collection}
J_\mathrm{e} = -e s_\mathrm{e} n_\mathrm{e} \sqrt{\frac{k_\mathrm{B}
T_\mathrm{e}}{2\pi m_\mathrm{e}}}\left( 1 + \frac{e\phi}{k_\mathrm{B}
T_\mathrm{e}} \right),
\end{equation}
where we use a sticking coefficient $s_\mathrm{e}=0.5$
\citep{draine78}. Given the low temperature of the gas, the incident
electrons have energies $\ll 1$\,eV, not enough to make secondary
electron emission important (e.g., \citealt{mendis94}). The ion
counterpart to equation~(\ref{eq:J_e_collection}) is found to be
negligible, since these particles are strongly repelled when $e\phi>0$
(e.g., \citealt{spitzer41}). Provided that relative drift velocities
between dust grains and electrons are smaller than the electron
thermal velocity (i.e., $v \lesssim 40$\,km\,s$^{-1}$), it is not
necessary to correct equation~(\ref{eq:J_e_collection}) for this
effect \citep{northrop96}, since positive charge is not provided by
ions but by photoelectrons. We set $T_\mathrm{e}$ equal to the gas
temperature $T$, equation~(\ref{eq:T_mean}).

The equilibrium grain potential is found by solving $J_\mathrm{ph} -
J_\mathrm{e} = 0$, the result being shown in Figure~\ref{fig:y},
normalized by the local value of $k_\mathrm{B}T$. As mentioned in the
main text, the contribution of the ejected photoelectrons to the
ambient electron density is very small. We estimate an upper limit to
the density of ejected photoelectrons as
\begin{equation}
n_\mathrm{e,dust} \le \left(\frac{e\phi}{k_\mathrm{B}T}\right)\frac{ak_\mathrm{B}T}{e^2 \lambda_\mathrm{D}^3}\sim 0.2\textrm{\,cm}^{-3}\ll n_\mathrm{e} 
\end{equation}
for $e\phi/(k_\mathrm{B}T) = 1\,000$, $T = 60$\,K, $n_\mathrm{e} =
4$\,cm$^{-3}$, and $a = 1$\,$\mu$m.  In the following, we provide a
rough estimate for the equilibrium potential, and show that it depends
strongly on the grain work function ($W$), but logarithmically on all
other parameters, making our results quite robust.

In the Wien regime, photon flux can be approximated as
\begin{equation}
F_\nu = f_0\left(\frac{\mathrm{AU}}{r}\right)^2 \nu^3 \exp{\left(-\frac{h \nu}{k_\mathrm{B} T_\star}\right)},
\label{eq:fnu}
\end{equation}
where $k_\mathrm{B} T_\star \approx 0.7$\,eV for $\beta$\,Pic, $r$ is
the distance to the star, and $f_0 =
F_\nu(\nu_0)\exp{(h\nu_0/[k_\mathrm{B}T_\star])}/\nu_0^3$, with
$F_\nu(\nu_0) \approx 3\times
10^{-9}$\,erg\,cm$^{-2}$\,s$^{-1}$\,Hz$^{-1}$ and $h\nu_0 = 4.4$\,eV.
We integrate the photoelectric current (equation \ref{eq:J_ph}) taking
only account of the exponential drop-off in the flux (and disregarding
factors of order unity outside the exponential factor),
\begin{equation}
J_\mathrm{ph} \sim \frac{e}{8}\,
\left[ \left(\frac{\mathrm{AU}}{r}\right)^2 \frac{f_0 \nu^3}{h}\frac{k_\mathrm{B}T_\star}{h\nu}
\right]_{h\nu \approx e\phi + W}\, \exp{\left( -\frac{e\phi + W}{k_\mathrm{B}T_\star}\right)} \sim
\left[ \left(\frac{\mathrm{AU}}{r}\right)^2 \frac{f_0 W^2 k_\mathrm{B}T_\star}{h^4}\right]\, 
\exp{\left( -\frac{e\phi + W}{k_\mathrm{B}T_\star}\right)}
\end{equation}
where we have used the fact that $W \geq e\phi$. Even if this may
not be a good approximation, the current depends exponentially on the
ratio $(e\phi+W)/k_\mathrm{B} T_\star$ and other factors outside the
exponential matter little. We also simplify the equation
(\ref{eq:J_e_collection}) by assuming $e\phi\sim k_\mathrm{B} T_\star
\gg k_\mathrm{B} T$. We equate the above expression to the photoelectric
current, and solve for the equilibrium potential:
\begin{equation}
e\phi = k_\mathrm{B} T_\star\ln \left[ \frac{1}{8s_\mathrm{e}n_\mathrm{e}} \left(\frac{\mathrm{AU}}{r}\right)^2
\left(\frac{f_0 W^2 k_\mathrm{B}T_\star}{h^4}\right)\,
\sqrt{\frac{2 \pi m_e}{k_\mathrm{B} T}}\,{T \over T_\star}\right] - W
\label{eq:logarithmic}
\end{equation}
This shows that the charging potential depends on all environment
variables logarithmically, and is only affected strongly by the work
function. Substituting in our fiducial values ($r = 100$~AU, $n_\mathrm{e} \sim
4$~cm$^{-3}$, $s_\mathrm{e} = 0.5$, $T = 60$~K) for the $\beta$~Pic disk, we
obtain $e\phi \sim 2.8$~eV for silicate dust ($W = 8$~eV) and $e\phi
\sim 5.5$~eV for carbon dust ($W = 4.4$~eV). Both values are much
greater than the values one expects if photoelectric charging is absent
($e\phi \sim k_{\mathrm{B}} T \sim 10^{-3}$~eV). We numerically
calculate these potentials (Fig. \ref{fig:y}) using a realistic stellar
spectrum and found them to be somewhat smaller than the values
obtained here.



\begin{thebibliography}{}
\bibitem[Artymowicz(1997)]{artymowicz97} Artymowicz, P. 1997, Annu. Rev. Earth Planet. Sci., 25, 175
\bibitem[Artymowicz et al.(1989)Artymowicz, Burrows \& Paresce]{artymowicz89} Artymowicz, P., Burrows, C., \& Paresce, F. 1989, \apj, 337, 494
\bibitem[Aumann et al.(1985)]{aumann85} Aumann, H.~H., et al. 1985, \apj, 278, L23
\bibitem[Backman \& Paresce(1993)]{backman93} Backman, D.~E., \& Paresce, F. 1993, in Protostars and Planets III, ed. E.~H. Levy, \&
J.~I. Lunine (Tucson: Univ. Arizona Press), 1253
\bibitem[Beust et al.(1989)]{beust89} Beust, H., Lagrange-Henri, A. M., Vidal-Madjar, A., \& Ferlet, R. 1989, \aap, 223, 304
\bibitem[Bodenheimer \& Lin(2002)]{bodenheimer_lin} Bodenheimer, P., \& Lin, D.~N.~C. 2002, Annu. Rev. Planet. Sci., 30, 113
\bibitem[Brandeker(2004)]{alexis_thesis} Brandeker, A. 2004, PhD Thesis, Stockholm University, available from the author on request
\bibitem[Brandeker et al.(2004)]{brandeker04} Brandeker, A., Liseau, R., Olofsson, G., \& Fridlund, M. 2004, \aap, 413, 681 
\bibitem[Chandrasekhar(1941)]{chandra41} Chandrasekhar, S. 1941, \apj, 93, 285
\bibitem[Chandrasekhar(1943)]{chandra43} Chandrasekhar, S. 1943, \apj, 97, 255
\bibitem[Crifo et al.(1997)]{crifo97} Crifo, F., Vidal-Madjar, A., Lallement, R., Ferlet, R., \& Gerbaldi, M. 1997, \aap, 320, L29
\bibitem[Draine(1978)]{draine78} Draine, B.~T. 1978, \apjs, 36, 595
\bibitem[Ferland et al.(1998)]{ferland98} Ferland, G.~J., Korista, K.~T., Verner, D.~A., Ferguson, J.~W., Kingdon, J.~B., \& Verner, E.~M. 1998, \pasp, 110, 761
\bibitem[Ferlet et al.(1987)]{ferlet87} Ferlet, R., Hobbs, L.~M., \& Vidal-Madjar, A. 1987, \aap, 185, 267
\bibitem[Freudling et al.(1995)]{freudling95} Freudling, W., Lagrange, A.-M., Vidal-Madjar, A., Ferlet, R., \& Forveille, T. 1995, \aap, 301, 231
\bibitem[Gray(1976)]{gray76} Gray, D.~F. 1976, The Observation and Analysis of Stellar Photospheres, (New York: Wiley)
\bibitem[Greenberg(1971)]{greenberg71} Greenberg, J.~M. 1971, \aap, 12, 240
\bibitem[Hauschildt et al.(1999)Hauschildt, Allard \& Baron]{hauschildt99} Hauschildt, P.~H., Allard, F., \& Baron, E. 1999, \apj, 512, 377
\bibitem[Heap et al.(2000)]{heap00} Heap, S.~R., Lindler, D.~J., Lanz, T.~M., Cornett, R.~H., Hubeny, I., Maran, S.~P., \& Woodgate, B.
2000, \apj, 539, 435
\bibitem[Hilborn(1982)]{hilborn82} Hilborn, R.~C. 1982, Am. J. Phys., 50, 982
\bibitem[Hobbs et al.(1985)]{hobbs85} Hobbs, L.~M., Vidal-Madjar, A., Ferlet, R., Albert, C.~E., \& Gry, C. 1985, \apj, 293, L29
\bibitem[H\o g et al.(2000)]{hog00} H\o g, E., et al. 2000, \aap, 335, L27
\bibitem[Hor\'anyi(1996)]{horanyi96} Hor\'anyi, M. 1996, \araa, 34, 383
\bibitem[Johnson(2005)]{cccbdb} Johnson, R.~D. (ed.) 2005, NIST Computational Chemistry Comparison and Benchmark Database,
NIST Standard Reference Database Number 101, (Gaithersburg, MD: NIST), Release 11, http://srdata.nist.gov/cccbdb
\bibitem[Kamp \& van Zadelhoff(2001)]{kamp01} Kamp, I., \& Zadelhoff, G.-J. 2001, \aap, 373, 641 
\bibitem[Keller et al.(1994)]{keller94} Keller, L.~P., Thomas, K.~L., \& McKay, D.~S. in Carbon in Primitive
Interplanetary Dust Particles, ed. E.~Zolensky, T.~L. Wilson, F.~J.~M. Rietmeijer, \& G.~J. Flynn (New York: AIP), 159
\bibitem[Lagage \& Pantin(1994)]{lagage94} Lagage, P.~O., \& Pantin, E. 1994, \nat, 369, 628
\bibitem[Lagrange-Henri et al.(1988)]{lagrange88} Lagrange-Henri, A.~M., Vidal-Madjar, A., \& Ferlet, R. 1988, \aap, 190, 275
\bibitem[Lagrange et al.(1995)]{lagrange95} Lagrange, A.-M., Vidal-Madjar, A., Deleuil, M., Emerich, C., Beust, H., \& Ferlet, R. 
1995, \aap, 296, 499 
\bibitem[Lagrange et al.(1996)]{lagrange96} Lagrange, A.-M., et al. 1996, \aap, 310, 547
\bibitem[Lagrange et al.(1998)]{lagrange98} Lagrange, A.-M., et al. 1998, \aap, 330, 1091
\bibitem[Lagrange et al.(2000)Lagrange, Backman \& Atymowicz]{lagrange2000} Lagrange, A.-M., Backman, D.~E., \& Artymowicz, P. 2000, 
in Protostars and Planets IV, ed. V.~Mannings, A.~P.~Boss, \& S.~S.~Russell (Tuczon: Univ. Arizona Press), 639
\bibitem[Lecavelier des Etangs et al.(2001)]{lecavelier01} Lecavelier des Etangs, A., et al. 2001, \nat, 412, 706
\bibitem[Liseau(2003)]{liseau03} Liseau, R. 2003, in Towards Other Earths: DARWIN/TPF and the Search for Extrasolar Terrestrial Planets,
ed. M.~Fridlund, T.~Henning, \& H.~Lacoste (Noordwijk: ESA Publ. Division), 135
\bibitem[Martin et al.(1999)]{asd} Martin, W.~C., Fuhr, J.~R., Kelleher, D.~E., Musgrove, A., Sugar, J., Wiese, W.~L., Mohr, P.~J., \& Olsen, K. 1999, 
NIST Atomic Spectra Database, (Gaithersburg, MD: NIST), version 2.0, http://physics.nist.gov/asd2
\bibitem[Mendis \& Rosenberg(1994)]{mendis94} Mendis, D.~A., \& Rosenberg, M. 1994, \araa, 32, 419
\bibitem[Northrop \& Birmingham(1990)]{northrop90} Northrop, T.~G., \& Birmingham, T.~J. 1990, Planet.~Space~Sci., 38, 319 
\bibitem[Northrop \& Birmingham(1996)]{northrop96} Northrop, T.~G., \& Birmingham, T.~J. 1996, J.~Geophys.~Res., 101, 793
\bibitem[Olofsson et al.(2001)]{olofsson01} Olofsson, G., Liseau, R., \& Brandeker, A. 2001, \apj, 563, L80
\bibitem[Rafikov(2004)]{rafikov04} Rafikov, R.~R. 2004, \aj, 128, 1348
\bibitem[Roberge et al.(2005)]{roberge05} Roberge, A., Weinberger, A.~J., Feldman, P.D., Deleuil, M., \& Bouret, J.-C. 2005, 
poster \#8036 at Protostars and Planets V, Hawaii, USA
\bibitem[Royer et al.(2002)]{royer02} Royer, F., Gerbaldi, M., Faraggiana R., G\'omez, A.~E. 2002, \aap, 381, 105
\bibitem[Shu et al.(1987)Shu, Adams \& Lizano]{shuadamslizano} Shu, F.~H., Adams, F.~C., \& Lizano, S. 1987, \araa, 25, 23
\bibitem[Smith \& Terrile(1984)]{smith84} Smith, B.~A., \& Terrile, R.~J. 1984, Sci, 226, 1421
\bibitem[Spitzer(1941)]{spitzer41} Spitzer, L. 1941, \apj, 93, 369
\bibitem[Spitzer(1956)]{spitzer56} Spitzer, L. 1956, Physics of Fully Ionized Gases (New York: Interscience)
\bibitem[Spitzer(1978)]{spitzer78} Spitzer, L. 1978, Physical Processes in the Interstellar Medium (New York: Wiley)
\bibitem[Th\'ebault et al.(2003)]{thebault03} Th\'ebault, P., Augereau, J.~C., \& Beust, H. 2003, \aap, 408, 775
\bibitem[Th\'ebault \& Augereau(2005)]{thebault05} Th\'ebault, P., \& Augereau, J.~C. 2005, \aap, 437, 141
\bibitem[Thi et al.(2001)]{thi01} Thi, W.~F., et al. 2001, \nat, 409, 60
\bibitem[Weinberger et al. (2003)]{weinberger} Weinberger, A. J., Becklin, E. E., Zuckerman, B. 2003, \apj, 584, 33
\bibitem[Weingartner \& Draine(2001)]{weingartner01} Weingartner, J.~C., \& Draine, B.~T. 2001, \apj, 134, 263
\bibitem[Zuckerman et al.(2001)]{zuckerman01} Zuckerman, B., Song, I., Bessel, M.~S., \& Webb, R.~A. 2001, \apj, 562, L87 
\end{thebibliography}
\end{document}